\documentclass[twocolumn]{aastex631}
\UseRawInputEncoding
\usepackage{natbib}
\usepackage{graphicx}
\usepackage{scrextend}
\usepackage[caption=false]{subfig}
\graphicspath{{figures/}}
\usepackage{float}
\usepackage{footnote}
\usepackage{booktabs,threeparttable}

\newcommand{\oh}{$12+\log{\mathrm{O/H}}$}
\newcommand{\Te}{T$_{\mathrm e}$}

\newcommand{\Mstar}{\mathrm{M}_*}
\newcommand{\Msun}{\mathrm{M}_{\sun}}
\newcommand{\Msunyr}{\mathrm{M}_{\sun}~\mathrm{yr}^{-1}}

\newcommand{\Hii}{H~{\sc ii}}

\newcommand{\Oii}{[O~{\sc ii}]}
\newcommand{\Oiii}{[O~{\sc iii}]}
\newcommand{\Oiiit}{[O~{\sc iii}]$\lambda$4363}
\newcommand{\Neiii}{[Ne~{\sc iii}]}

\newcommand{\Ha}{H$\alpha$}
\newcommand{\Hb}{H$\beta$}

\newcommand{\Nii}{[N~{\sc ii}]}
\newcommand{\Sii}{[S~{\sc ii}]}
\newcommand{\Siii}{[S~{\sc iii}]}
\newcommand{\SiiHa}{\Sii/(\Ha+\Nii)}
\newcommand{\SigHa}{$\Sigma_{H\alpha}$}

\shorttitle{Understanding high-redshift BPT offsets with spatially resolved spectroscopy}
\shortauthors{Hirtenstein et al.}

\begin{document}

\title{Disentangling the physical origin of emission line ratio offsets at high redshift with spatially resolved spectroscopy}

\email{jhirtenstein@ucdavis.edu}

\author[0000-0002-5368-8262]{Jessie Hirtenstein\affiliation{}}
\author[0000-0001-5860-3419]{Tucker Jones\affiliation{}}
\author[0000-0003-4792-9119]{Ryan L. Sanders\affiliation{}}
\altaffiliation{NHFP Hubble Fellow}
\affil{Department of Physics and Astronomy, University of California, Davis, CA, USA}
\author{Crystal L. Martin\affiliation{}}
\affil{Department of Physics, University of California, Santa Barbara, Santa Barbara, CA, USA}
\author{M. C. Cooper\affiliation{}}
\affil{Department of Physics and Astronomy, University of California, Irvine, Irvine CA, USA}
\author{Gabriel Brammer\affiliation{}}
\affil{Cosmic Dawn Centre, Niels Bohr Institute, University of Copenhagen, Copenhagen, Denmark}
\author{Tommaso Treu\affiliation{}}
\affil{Department of Physics and Astronomy, University of California, Los Angeles, CA, USA}
\author{Kasper Schmidt\affiliation{}}
\affil{Leibniz-Institut für Astrophysik Potsdam (AIP), Potsdam, Germany}
\author{Alice Shapley\affiliation{}}
\affil{Department of Physics and Astronomy, University of California, Los Angeles, CA, USA}



\begin{abstract}
We present spatially resolved Hubble Space Telescope grism spectroscopy of 15 galaxies at $z\sim0.8$ drawn from the DEEP2 survey. We analyze \Ha+\Nii, \Sii\ and \Siii\ emission on kpc scales to explore which mechanisms are powering emission lines at high redshifts, testing which processes may be responsible for the well-known offset of high redshift galaxies from the $z\sim0$ locus in the \Oiii/\Hb\ versus \Nii/\Ha\ BPT (Baldwin-Phillips-Terlevich) excitation diagram. 
We study spatially resolved emission line maps to examine evidence for active galactic nuclei (AGN), shocks, diffuse ionized gas (DIG), or escaping ionizing radiation, all of which may contribute to the BPT offsets observed in our sample. We do not find significant evidence of AGN in our sample and quantify that, on average, AGN would need to contribute $\sim$25\% of the \Ha\ flux in the central resolution element in order to cause the observed BPT offsets. We find weak ($2\sigma$) evidence of DIG emission at low surface brightnesses, yielding an implied total DIG emission fraction of $\sim$20\%, which is not significant enough to be the dominant emission line driver in our sample. In general we find that the observed emission is dominated by star forming \Hii\ regions. We discuss trends with demographic properties and the possible role of $\alpha$-enhanced abundance patterns in the emission spectra of high redshift galaxies. Our results indicate that photo-ionization modeling with stellar population synthesis inputs is a valid tool to explore the specific star formation properties which may cause BPT offsets, to be explored in future work.

\end{abstract}



\section{Introduction}
\label{sec:intro}

Much of our knowledge of galaxy evolution across cosmic time is based on strong nebular emission lines at rest-frame optical wavelengths. Analysis of nebular line emission reveals properties of star formation and ionized gas including the total star formation rate (SFR), metallicity, gas density, and ionization parameter. Such studies have proven highly productive, with the advent of sensitive multiplexed spectrographs enabling comprehensive emission line studies at high redshifts \citep[e.g.,][]{Steidel_2014,Shapley_2015,Momcheva_2016,Sanders_2016a,Kashino_2019}. 

A key limitation of evolutionary studies is that some properties are typically not measured directly, in particular the gas metallicity (for which we will adopt the standard convention in terms of oxygen abundance: \oh). Instead, nearly all studies at high redshift rely on calibrations of strong emission line ratios to estimate the galaxy metallicity (but see, e.g., \citealt{Sanders_2016b,Sanders_2020b,Patricio_2018,Gburek_2019} for small $z>1$ samples that have the direct metallicities based on electron temperature T$_e$). When applying these strong-line diagnostics, it is implicitly assumed that the high redshift samples have similar interstellar medium (ISM) physical conditions as star forming regions in nearby ($z\sim0$) galaxies, which are used to calibrate these methods. 

Large spectroscopic samples of $z>1$ galaxies have revealed significant redshift evolution in the excitation sequences of strong emission line ratios. Most notably, it is now well-established that there is an offset in the ``BPT diagram'' of \Oiii/\Hb\ versus \Nii/\Ha\ \citep{BPT_1981} between $z>1$ samples and the locus of $z\sim0$ star-forming galaxies \citep[e.g.,][]{Steidel_2014,Shapley_2015,Kashino_2017}.
This BPT offset suggests an evolution in the underlying physical properties governing emission line production, such as the shape of the ionizing spectrum, gas density, ionization parameter, and chemical abundance pattern \citep[e.g.,][]{Kewley_2013a}.
One consequence of changing ISM conditions is that the translation between strong-line ratios and metallicity will also evolve, such that applying locally calibrated diagnostics at high redshifts will yield systematically biased metallicity estimates \citep[e.g.,][]{Bian_2018,Patricio_2018,Sanders_2020c}.
This systematic error will directly affect inferences on galaxy formation and evolution based on the observed chemical evolution of galaxies \citep[e.g.,][]{Lilly_2013,Zahid_2014,Troncoso_2014,Sanders_2020c}. 

Several hypotheses for the observed emission-line ratio offsets at high redshift have been suggested (e.g. \citealt{Kewley_2013a}). These include emission mixing from low-luminosity active galactic nuclei (AGN; e.g., \citealt{Wright_2010}) or shocked gas (e.g., \citealt{Yuan_2012}), contributions from diffuse ionized gas (DIG; e.g.,  \citealt{Zhang_2016}), an enhanced N/O abundance ratio (e.g., \citealt{Masters_2014, Shapley_2015, Masters_2016}), high ionization parameter (e.g., \citealt{Kewley_2013b}), and hard ionizing spectra driven by super-solar $\alpha$/Fe abundance patterns \cite[e.g.,][]{Steidel_2016, Shapley_2019, Sanders_2020b, Topping_2020a, Topping_2020b}. 
To distinguish among these possibilities -- and thereby recover accurate results with emission line studies -- we require sensitive physical measurements beyond those typically available from the strongest nebular lines.

A critical step forward is the measurement of gas-phase metallicity and temperature using the ``direct'' \Te\ method based on auroral emission lines. In our earlier work, \cite{Jones_2015} presented \Te-based abundance measurements for a sample of star forming galaxies at $z\sim0.8$ drawn from the DEEP2 Galaxy Redshift Survey (DEEP2; \citealt{Davis_2003, Newman_2013}). 
\cite{Jones_2015} found that the relations between direct metallicity and rest-optical line ratios of \Oii, \Oiii, \Hb, and \Neiii\ for the $z\sim0.8$ sample are remarkably consistent with the sample of $z\sim0$ galaxies from \citet{Izotov_2006}.
While this comparison appeared to suggest no evolution in metallicity calibrations over $z=0-0.8$, a key caveat is that the \citet{Izotov_2006} sample comprises extreme local galaxies with high specific SFRs analogous to those of high-redshift galaxies due to selection effects, as noted by \citet{Sanders_2020b,Sanders_2020c}.

Comparison to a wider range of local galaxy properties is important for fully understanding the evolution of metallicity calibrations. 
Indeed, in comparing more representative samples of local galaxies and \Hii\ regions to \Te-based samples at higher redshifts, \cite{Sanders_2020b} found evidence that strong-line metallicity calibrations evolve between $z\sim0$ and $z\sim1.5-3.5$. 
By modeling the spectra of their $z\sim2$ \Te\ sample, these authors found that the evolution of strong-line calibrations is due to a harder ionizing spectrum driven by $\alpha$-enhancement of young, massive stars, in agreement with strong-line and rest-UV studies providing similar explanations for the $z\sim2$ BPT diagram offset \citep{Steidel_2016,Strom_2018,Shapley_2019,Sanders_2020a,Topping_2020a,Topping_2020b, Runco_2021}. 
While these studies based on integrated galaxy spectra have yielded great progress in understanding the drivers of galaxy line ratio evolution, there remain potential contributors that cannot be easily identified in integrated measurements including DIG emission, weak AGN, and shocks.

\begin{figure*}
\centering

\subfloat[MaNGA galaxy 10001-6102 with central region dominated by Seyfert type AGN emission. Since the AGN-dominated nucleus (pink) has the same range of {\SiiHa}) as the purely star formation spaxels (teal), we cannot use this diagnostic to discern between Seyfert AGN and star forming regions.]
{\includegraphics[width= 0.9\textwidth]{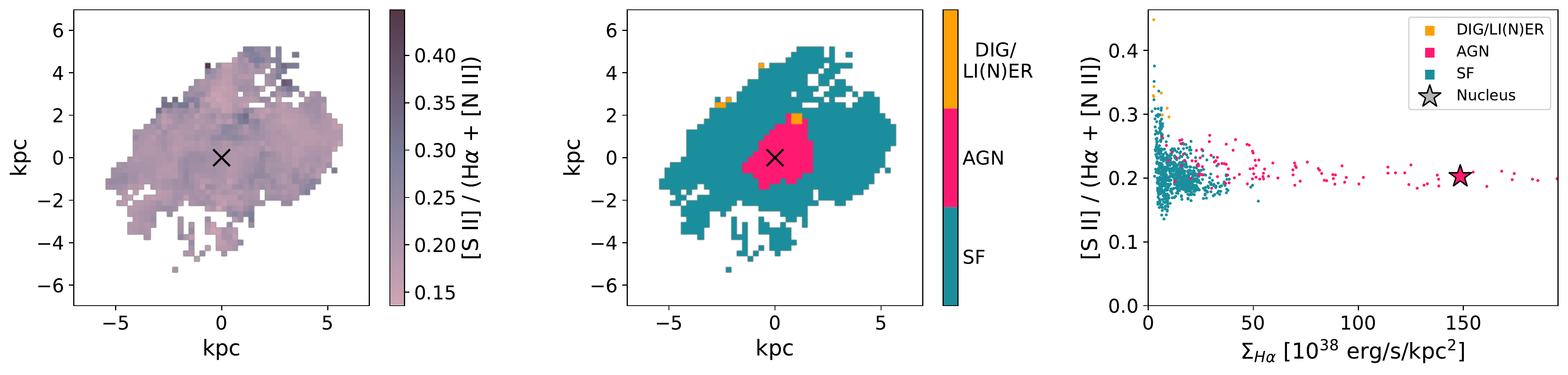}
\label{fig:MANGA_seyfert}}\\

\subfloat[MaNGA galaxy 8134-9102 with central region dominated by LINER type AGN emission. The spaxels separate into two clear populations, notably those categorized as LI(N)ER emission (orange) are found in the nuclear regions and have elevated levels of {\SiiHa} across all surface brightnesses relative to the purely star forming regions (teal).]
{\includegraphics[width=0.9\textwidth]{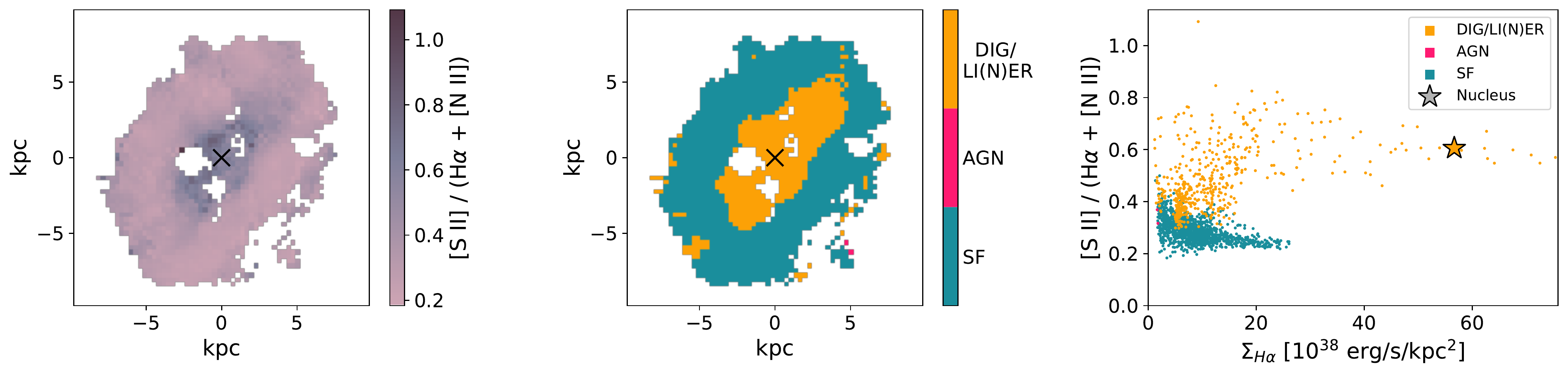}
\label{fig:MANGA_liner}}\\

\subfloat[MaNGA galaxy 8082-12702 with contributions from shocks and DIG. The spaxels categorized as DIG emission (orange) have elevated {\SiiHa} relative to star forming regions (teal) and are concentrated in the lowest surface brightness regions.]
{\includegraphics[width=0.9\textwidth]{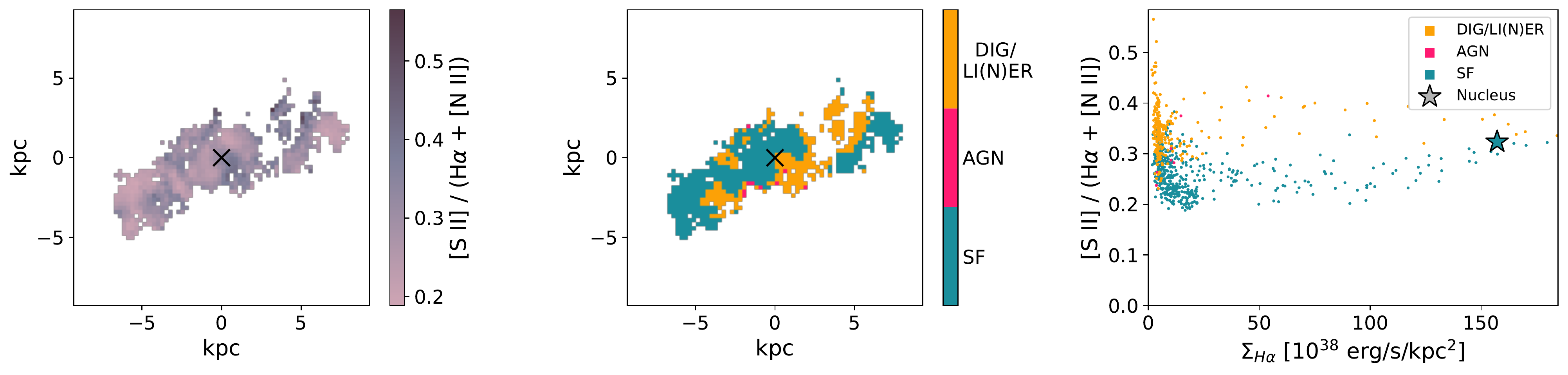}
\label{fig:MANGA_dig_lier}}\\

\subfloat[MaNGA galaxy 8085-12704 with emission lines dominated by star formation (teal) in all spatial regions. We see only one population in the rightmost panel, where the ratio of {\SiiHa})  remains consistently flat for all spaxels across all surface brightnesses.]
{\includegraphics[width=0.9\textwidth]{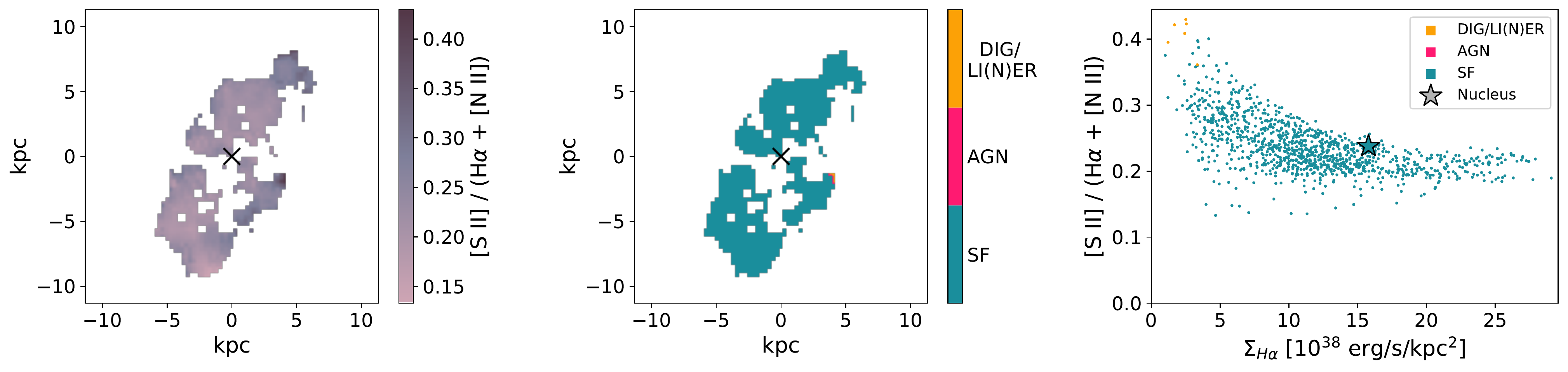}
\label{fig:MANGA_SF}}\\
\caption{Demonstration of the methods applied in this paper to distinguish emission line excitation from different sources. This figure shows examples of MaNGA galaxies showing a range of excitation mechanisms including regions dominated by (a) Seyfert AGN, (b) LINER type AGN, (c) shocks or DIG and (d) star formation. \textit{Left panels}: Spatially resolved \SiiHa\ maps. Black crosses denote galaxy nuclei. \textit{Middle panels}: Spaxels color coded by dominant forms of emission following the demarcations of \cite{Kewley_2001, Kewley_2006}: DIG/LI(N)ER (orange), AGN (pink) and star formation (teal). \textit{Right panels}: \SiiHa\ plotted against \Ha+\Nii\ surface brightness, following the same dominant emission color codings. These galaxies show how different emission regions separate into clear populations when analyzed with this metric of spatially resolved \SiiHa, which we apply to HST grism data of $z\simeq0.8$ galaxies in this work.}
\label{fig:MANGA}
\end{figure*}

This work is concerned with the spatially resolved analysis of a subset of galaxies studied by \cite{Jones_2015} to further understand the ionizing sources that power nebular line emission in high-z galaxies. Here we use \textit{Hubble Space Telescope} (HST) WFC3-IR grism spectroscopy covering the \Ha+\Nii\footnote{\Ha\ and \Nii$\lambda\lambda$6548,6584 emission lines are blended at the spectral resolution of the WFC3-IR data; we refer to the sum of these lines as \Ha+\Nii. Likewise \Sii\ refers to the sum of the blended \Sii$\lambda\lambda$6716,6731 lines.}, \Sii, and \Siii\ emission lines for 15 galaxies at $\simeq1$~kpc spatial resolution. Spatial mapping of these emission lines enables several key tests of the ionization mechanisms. 
In particular we seek to distinguish emission powered by star formation (i.e. \Hii\ regions), AGN, shocked gas and DIG. 
If AGN (particularly LINERs: low-ionization nuclear emission line regions) are present and responsible for BPT offsets, we expect to observe an elevated \SiiHa\ ratio in the central $\sim$1 kpc resolution element. Conversely if BPT offsets are due to shocked gas or DIG, we would expect elevated \SiiHa\ in low surface brightess regions. Contributions from density-bounded \Hii\ regions with escaping ionizing radiation are expected to result in elevated \Siii/\Sii\ ratios in the corresponding spatial regions. 
The diffraction-limited \textit{HST} spectroscopic data presented herein therefore allow us to differentiate between these possibilities and discern the origin of redshift evolution in the BPT diagram. Understanding the cause of these line ratio offsets will allow us to better calibrate metallicity measurements between the low and high redshift universe. 

This paper is structured as follows. In Section~\ref{sec:manga}, we use spatially resolved spectroscopy of example galaxies at $z\sim0$ to verify the methodology that we will apply to \textit{HST}/WFC3-IR grism spectra. We describe our \textit{HST} sample and data reduction methods in Section~\ref{sec:HST}. In Section~\ref{sec:analysis} we describe our analysis techniques and discuss one sample object in depth. In Sections~\ref{sec:discussion} and \ref{sec:SF} we analyze the various processes which can potentially power the observed emission lines. Finally, in Section \ref{sec:summary}, we summarize our results and discuss implications of this work for future research. 
Where necessary we assume a flat $\Lambda$CDM cosmology with H$_0$ = 69.6 km s$^{-1}$ Mpc$^{-1}$, $\Omega_m$ = 0.286, and $\Omega_{\Lambda}$ = 0.714. All magnitudes are on the AB system.

\section{Methodology: Distinguishing Ionization Sources}
\label{sec:manga}

To establish our methodology, we first consider examples of different sources of emission line excitation observed in low-redshift galaxies from the SDSS-IV Mapping Nearby Galaxies at APO (MaNGA) survey \citep{Bundy_2015}, which obtained optical integral field spectroscopy (IFS) at $\sim1$~kpc resolution for $\sim10,000$ low-redshift galaxies.
In this section we use MaNGA IFS data to demonstrate how various ionization scenarios can be clearly distinguished using the diagnostics available with HST grism data. Our primary diagnostic is the line ratio \SiiHa, combined with the \Ha+\Nii\ surface brightness and spatial location within a galaxy. 

We utilized the MaNGA catalogs and emission-line flux maps produced using the PIPE3D pipeline\footnote{Available online at https://www.sdss.org/dr14/manga/manga-data/manga-pipe3d-value-added-catalog/.} \citep{Sanchez_2016a,Sanchez_2016b}.
For each galaxy, we selected all spaxels with S/N$\ge$3 in \Ha, \Hb, \Oiii$\lambda$5007, \Nii$\lambda$6584, \Sii$\lambda$6716, and \Sii$\lambda$6731.
We then calculated the line ratios \Oiii/\Hb, \Nii/\Ha, \Sii/\Ha, and \SiiHa, noting that reddening correction is not required due to the close proximity of the wavelengths of the lines in each ratio.
Following \citet{Belfiore_2016}, we classified spaxels as primarily ionized by star formation (SF), Seyfert-like AGN, or as low-ionization emission regions (LIERs), based on cuts in the \Oiii/\Hb\ vs.\ \Sii/\Ha\ diagram using the demarcations of \citet{Kewley_2001,Kewley_2006}.
LIERs that are nuclear (e.g., LINERs) can be attributed to low-luminosity AGN, while LIERs that are extended throughout the disk are thought to originate from shocked gas and/or DIG emission \citep{Belfiore_2016}. 
We furthermore computed the observed \Ha+\Nii\ luminosity surface brightness (\SigHa) using the \Ha+\Nii\ fluxes, spaxel sizes, and measured redshifts.

We produced maps of \SiiHa\ and SF vs.\ LIER vs.\ AGN classification, as well as plots of \SiiHa\ vs. \SigHa\ (where \SigHa\ represents the \Ha+\Nii\ surface brightness) for many MaNGA targets in order to evaluate how well these observables can distinguish between the various ionization mechanisms
that we will use for our subsequent \text{HST} analysis.
Illustrative examples of four MaNGA targets are shown in Figure~\ref{fig:MANGA}.
In each row, the left panel displays a map of the \SiiHa\ ratio, the middle panel shows the ionization mechanism classification for each spaxel, and the right panel presents \SiiHa\ as a function of \SigHa, with the properties of the central 1~kpc indicated by a star.
For purposes of Figure~\ref{fig:MANGA}, we adopt a single classification for shocks, DIG, LIER, and LINER (i.e., ``DIG/LI(N)ER'') as these origins are not clearly distinguished by the emission line ratios alone. In our analysis herein we adopt the following conventions: concentrated nuclear LIER emission is considered a signature of {\it LINER AGN} (e.g., Figure~\ref{fig:MANGA_liner}), while spatially extended low-surface brightness LIER emission is considered a signature of {\it shocks} and {\it DIG} (e.g., Figure~\ref{fig:MANGA_dig_lier}). In our analysis, we distinguish between LINER and DIG/LIER regions by their spatial distribution and \Ha+\Nii\ surface brightnesses: LINER emission is nuclear in origin and thus will be concentrated in the central regions of the galaxies, typically with higher \Ha+\Nii\ surface brightness whereas DIG/LIER regions are located throughout the disk and in galaxy outskirts, in regions with predominantly lower \Ha+\Nii\ surface brightnesses.

  \begin{table*}
    \centering
    \footnotesize
    \begin{threeparttable}
      \caption{Target Properties}\label{tab:target_sample}
      \begin{tabular}{ c c c c c c c }
        \toprule
        Target ID & RA & Dec & Redshift & BPT offset\tnote{a} &  Magnitude & H$\alpha$ Equivalent Width\\
        \midrule
          &  &  & & dex & & \AA \\
        \bottomrule
	 DEEP2-13016475 & 14:20:57.8616 & +52:56:41.83 & 0.747 & 0.065 & 23.35 $\pm$ 0.04 & 1516.8 $\pm$ 93.7 \\
	 DEEP2-13043682 & 14:20:08.3688 & +53:06:37.57 & 0.761 & 0.097 & 22.31 $\pm$ 0.02 & 240.4 $\pm$ 26.6 \\
	 DEEP2-13043716 & 14:19:48.0672 & +53:05:12.62 & 0.781 & 0.111 & 21.11 $\pm$ 0.02 & 305.4 $\pm$ 15.7 \\
	 DEEP2-14018918 & 14:21:45.4104 & +53:23:52.71 & 0.770 & 0.070 & 22.97 $\pm$ 0.03 & 618.8 $\pm$ 39.2 \\
	 DEEP2-21021292 & 16:46:35.3952 & +34:50:27.80 & 0.763 & -0.001\tnote{b} & 22.14 $\pm$ 0.03 & 246.4 $\pm$ 11.9 \\
 	DEEP2-21027858 & 16:46:28.9848 & +34:55:19.82 & 0.841 & -0.115\tnote{b} & 22.83 $\pm$ 0.05 & 886.3 $\pm$ 51.8\tnote{c} \\
	 DEEP2-22020856 & 16:51:31.4496 & +34:53:15.82 & 0.796 & 0.092 & 23.02 $\pm$ 0.04 & 438.5 $\pm$ 31.3 \\
	 DEEP2-22028402 & 16:51:31.9272 & +34:55:29.28 & 0.797 & 0.076 & 20.28 $\pm$ 0.02 & 325.0 $\pm$ 4.9 \\
	 DEEP2-22022835 & 16:50:34.5960 & +34:52:52.09 & 0.843 & 0.001 & 23.50 $\pm$ 0.05 & 1035.1 $\pm$ 191.9\tnote{c} \\
	 DEEP2-22032252 & 16:53:03.4608 & +34:58:48.71 & 0.748 & 0.295 & 24.17 $\pm$ 0.07 & 700.0 $\pm$ 107.4 \\
 	DEEP2-22044304 & 16:51:20.3232 & +35:02:32.38 & 0.793 & 0.144 & 23.20 $\pm$ 0.04 & 990.0 $\pm$ 93.5 \\
	 DEEP2-41059446 & 02:26:21.4913 & +00:48:06.64 & 0.779 & 0.016 & 22.17 $\pm$ 0.02 & 344.0 $\pm$ 25.4 \\
	 DEEP2-42045870 & 02:30:10.6027 & +00:41:17.37 & 0.835 & -0.030\tnote{b} & 21.94 $\pm$ 0.02 & 396.6 $\pm$ 40.0 \\
	 DEEP2-31046514 & 23:27:07.5024 & +00:17:41.18 & 0.790 & 0.058 & 22.85 $\pm$ 0.03 & 529.9 $\pm$ 43.7 \\
	 DEEP2-31047144 & 23:26:55.4256 & +00:17:52.60 & 0.855 & 0.182 & 22.90 $\pm$ 0.03 & 848.5 $\pm$ 44.9 \\
      \end{tabular}
      \begin{tablenotes}
        \item[a]BPT offsets are calculated as the minimum distance from the $z\sim0$ star forming locus on the \Oiii/\Hb\ versus \Nii/\Ha\ BPT diagram.
        \item[b]These targets fall beneath the star forming locus in \Oiii/\Hb\, so we list their BPT offsets as negative.
        \item[c]These objects have negative \Nii\ best-fit fluxes, so their equivalent widths assume no contribution of \Nii\ to the blended \Ha+\Nii\ signal.
      \end{tablenotes}
    \end{threeparttable}
  \end{table*}

\subsection{Weak AGN}

A combination of AGN and star formation excitation can naturally explain the offsets in the BPT diagram observed at high redshift. This scenario can be distinguished by spatially resolving the nuclear emission line ratios from those of surrounding star forming regions \citep[as demonstrated by, e.g.,][]{Wright_2010}. Figures \ref{fig:MANGA_seyfert} and \ref{fig:MANGA_liner} show galaxies with Seyfert and LINER AGN, respectively, in the central regions. In both cases the galaxy outskirts show significant emission from star forming regions, leading to composite integrated spectra. 
As seen in the right-hand panel of Figure~\ref{fig:MANGA_seyfert}, we cannot use \SiiHa\ as a diagnostic to distinguish Seyfert AGN from star forming \Hii\ regions, because the AGN-dominated nuclei and \Hii\ regions both have the same range of \SiiHa\ flux ratios.
However we can see in Figure~\ref{fig:MANGA_liner} that LINER-like low-luminosity AGN are clearly distinguished as having a much higher central \SiiHa\ ($\simeq0.6$ in this case), compared to $\simeq0.2$--0.3 for \Hii\ regions (teal spaxels). 
Therefore, we expect to observe regions with elevated \SiiHa\ in the centers of our target galaxies (likely with high surface brightness) if they have significant emission from LINER AGN.

The differences in \SiiHa\ result from LINERs having characteristically higher \Sii/\Ha\ ratios than Seyferts, though comparable \Nii/\Ha. Both types of AGN generally have high ratios of \Nii/\Ha\ and \Sii/\Ha\ compared to \Hii\ regions. Although Seyfert AGN are not cleanly distinguished from \Hii\ regions using the \SiiHa\ diagnostic, we note that other signatures such as X-ray luminosity and broad emission lines can be used to differentiate these sources. 

  \begin{table*}
    \centering
    \footnotesize
    \begin{threeparttable}
      \caption{Emission Line Measurements}\label{tab:EL_fluxes}
      \begin{tabular}{c c c c c c}
        \toprule
        Target ID & integrated \Ha+\Nii\tnote{a} & integrated \Sii\tnote{a} & integrated \Siii\tnote{a} & Keck \Nii/\Ha\tnote{b} & Keck \Oiii/\Hb\tnote{c} \\
        \midrule
        & $10^{-16} erg/s/cm^{2} $ & $10^{-16} erg/s/cm^{2}$ & $10^{-16} erg/s/cm^{2}$ & & \\
        \bottomrule
         DEEP2-13016475 & 7.69 $\pm$ 0.16 & 0.68 $\pm$ 0.14 & 0.81 $\pm$ 0.09 & 0.019 $\pm$ 0.002 & 6.16 $\pm$ 0.06 \\
	 DEEP2-13043682 & 3.90 $\pm$ 0.21 & 1.00 $\pm$ 0.17 & 0.49 $\pm$ 0.17 & 0.062 $\pm$ 0.008 & 4.64 $\pm$ 0.08 \\
 	DEEP2-13043716 & 15.24 $\pm$ 0.29 & 1.78 $\pm$ 0.21 & 2.80 $\pm$ 0.29 & 0.147 $\pm$ 0.003 & 2.77 $\pm$ 0.05 \\
 	DEEP2-14018918 & 4.38 $\pm$ 0.20 & 0.46 $\pm$ 0.17 & 0.53 $\pm$ 0.21 & 0.021 $\pm$ 0.006 & 6.19 $\pm$ 0.08 \\
 	DEEP2-21021292 & 4.12 $\pm$ 0.20 & 0.71 $\pm$ 0.17 & 0.22 $\pm$ 0.18 & 0.055 $\pm$ 0.007 & 3.80 $\pm$ 0.07 \\
 	DEEP2-21027858 & 4.96 $\pm$ 0.13 & 0.65 $\pm$ 0.11 & 0.72 $\pm$ 0.30 & -0.027 $\pm$ 0.017\tnote{d} & 4.91 $\pm$ 0.11 \\
	 DEEP2-22020856 & 3.30 $\pm$ 0.17 & 0.41 $\pm$ 0.15 & 0.27 $\pm$ 0.24 & 0.045 $\pm$ 0.006 & 5.20 $\pm$ 0.08 \\
	 DEEP2-22028402 & 27.67 $\pm$ 0.31 & 3.68 $\pm$ 0.26 & 4.49 $\pm$ 0.35 & 0.147 $\pm$ 0.001 & 2.49 $\pm$ 0.03 \\
	 DEEP2-22022835 & 2.57 $\pm$ 0.15 & 0.08 $\pm$ 0.13 & 0.70 $\pm$ 0.34 & -0.003 $\pm$ 0.012\tnote{d} & 5.22 $\pm$ 0.12 \\
 	DEEP2-22032252 & 2.22 $\pm$ 0.16 & 0.33 $\pm$ 0.14 & 0.18 $\pm$ 0.09 & 0.123 $\pm$ 0.023 & 5.37 $\pm$ 0.11 \\
 	DEEP2-22044304 & 5.44 $\pm$ 0.17 & 0.85 $\pm$ 0.15 & 0.63 $\pm$ 0.25 & 0.054 $\pm$ 0.003 & 5.51 $\pm$ 0.04 \\
 	DEEP2-41059446 & 4.74 $\pm$ 0.31 & 1.24 $\pm$ 0.28 & 1.40 $\pm$ 0.38 & 0.031 $\pm$ 0.006 & 4.90 $\pm$ 0.13 \\
 	DEEP2-42045870 & 3.58 $\pm$ 0.25 & 0.82 $\pm$ 0.23 & 0.06 $\pm$ 0.47 & 0.036 $\pm$ 0.005 & 4.16 $\pm$ 0.07 \\
 	DEEP2-31046514 & 4.03  $\pm$ 0.24 & 1.01 $\pm$ 0.22 & 0.55 $\pm$ 0.33 & 0.043 $\pm$ 0.014 & 4.85 $\pm$ 0.12 \\
 	DEEP2-31047144 & 5.29 $\pm$ 0.15 & 0.56 $\pm$ 0.13 & 0.50 $\pm$ 0.58 & 0.046 $\pm$ 0.010 & 6.45 $\pm$ 0.13 \\
      \end{tabular}
      \begin{tablenotes}
        \item[a]Measured from HST WFC3-IR grism spectra.
        \item[b]\Nii$\lambda$6584/\Ha, measured from MOSFIRE or NIRSPEC spectra, as described in Section~\ref{sec:keck_spectra}.
        \item[c]Measured from DEEP2 spectra as described in \cite{Jones_2015}. Here \Oiii\ refers to the $\lambda$5007 line but is calculated as 2.98$\times$ the \Oiii$\lambda$4959 flux, as \Oiii$\lambda$5007 is redshifted beyond the spectral coverage in several targets. 
        \item[d]These objects have negative \Nii\ fluxes in their fits. Figures using these data show 2$\sigma$ upper limits.
      \end{tablenotes}
    \end{threeparttable}
  \end{table*}

\subsection{Shocks and Diffuse Ionized Gas}

As with LINER AGN activity, contributions from shocks or DIG can affect our observed emission line ratios. We can distinguish these effects via spatially resolved \SiiHa, where we expect elevated ratios to be localized just outside of the most luminous star forming regions and in galaxy outskirts in the case of shocks \citep[e.g.,][]{Newman_2012}, or in regions of lower \Ha\ surface brightness in the case of DIG \citep[e.g.,][]{Zhang_2016}.

The galaxy shown in Figure \ref{fig:MANGA_dig_lier} shows widespread regions dominated by DIG/shocked gas emission. These DIG, shock, or LIER-dominated areas can be distinguished from star forming \Hii\ regions, by their signature of elevated \SiiHa\ at relatively low \Ha+\Nii\ surface brightness $\Sigma_{H\alpha}$. Therefore, we expect to observe regions with elevated \SiiHa\ in the lowest surface brightness regions of our \textit{HST} galaxies if their emission is dominated by shocked gas or DIG.

\subsection{Escaping Ionizing Radiation}

The observed BPT offsets of high redshift galaxies can also be explained in part by a higher fraction of density-bounded \Hii\ regions, which may be expected at the corresponding higher star formation surface densities \citep[e.g.,][]{Beckman_2000,Alexandroff_2015}. If ionizing radiation escapes from these regions and into the surrounding intergalactic medium, we expect elevated \Siii/\Sii\ ratios in and around the density-bounded regions (which are likely the highest surface brightness regions; e.g., \citealt{Zastrow_2011}). 
For example, \cite{Zastrow_2013} show the presence of optically thin ``ionization cones'' characterized by strong \Siii\ emission extended over kpc scales, which reveal significant escaping ionizing radiation. 
HST grism spectroscopy presented in this work provides coverage of \Siii\ at $\sim$1 kpc resolution, although we generally do not detect the emission with high enough significance to search for such ionization cones, and thus we focus on other diagnostics. Nonetheless this approach may be of interest for galaxies with high confirmed or suspected ionizing escape fractions.

\subsection{Star Formation}

Finally, we consider the possibility that none of the above effects are significant, and that nebular emission in high-redshift galaxies is powered predominantly by star formation. In this case, offsets in the BPT diagram may be caused by different intrinsic properties of the massive stars and/or ionized gas in \Hii\ regions. 
If line emission in a galaxy is driven purely by star formation, we do not expect to see any extreme signatures with \SiiHa\ in relation to surface brightness or galactocentric radius. The relationship should remain a single population that is relatively flat with \SiiHa~$\lesssim$~0.3, as seen in Figure~\ref{fig:MANGA_SF}. Thus we do not expect to see strongly elevated \SiiHa\ at high or low surface brightnesses in our \textit{HST} sample if their emission is uniformly dominated by star formation.

\section{HST Sample Selection and Spectroscopy}
\label{sec:HST}

Having established the utility of spatially resolved line ratio mapping with $z\sim0$ galaxies, we will now apply the approach from Section~\ref{sec:manga} to HST grism spectroscopy at higher redshift ($z\sim0.8$). Despite the larger cosmological distance, HST delivers $\sim$1.0 kpc spatial resolution which is comparable to the low-redshift examples used to verify our methodology.

\begin{figure}
  \includegraphics[width=0.45\textwidth]{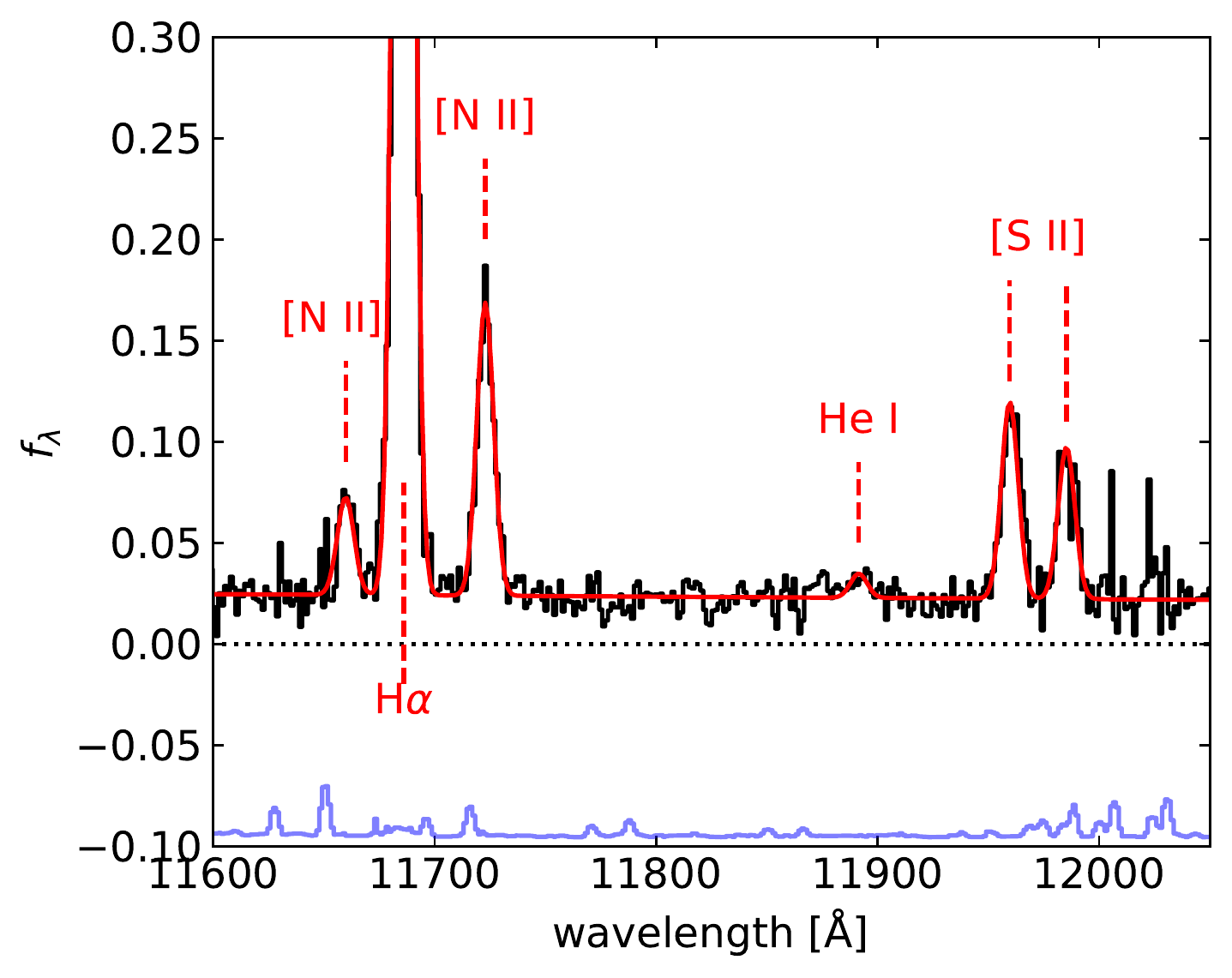}
  \caption{Near-IR Keck/MOSFIRE spectrum of DEEP2-13043716 at $z=0.78$ showing the \Ha, \Nii, and \Sii\ emission lines. The reduced spectrum is shown in black, normalized to have peak \Ha\ flux density $=1$. Gaussian fits to the strong emission lines are in red, and the 1-$\sigma$ error spectrum is in blue (offset by $-0.1$). The \Nii\ and \Ha\ lines, as well as the \Sii\ doublet are spectrally resolved. } 
  \label{fig:keck_obs}
\end{figure}

\begin{figure*}
  \includegraphics[width=\textwidth]{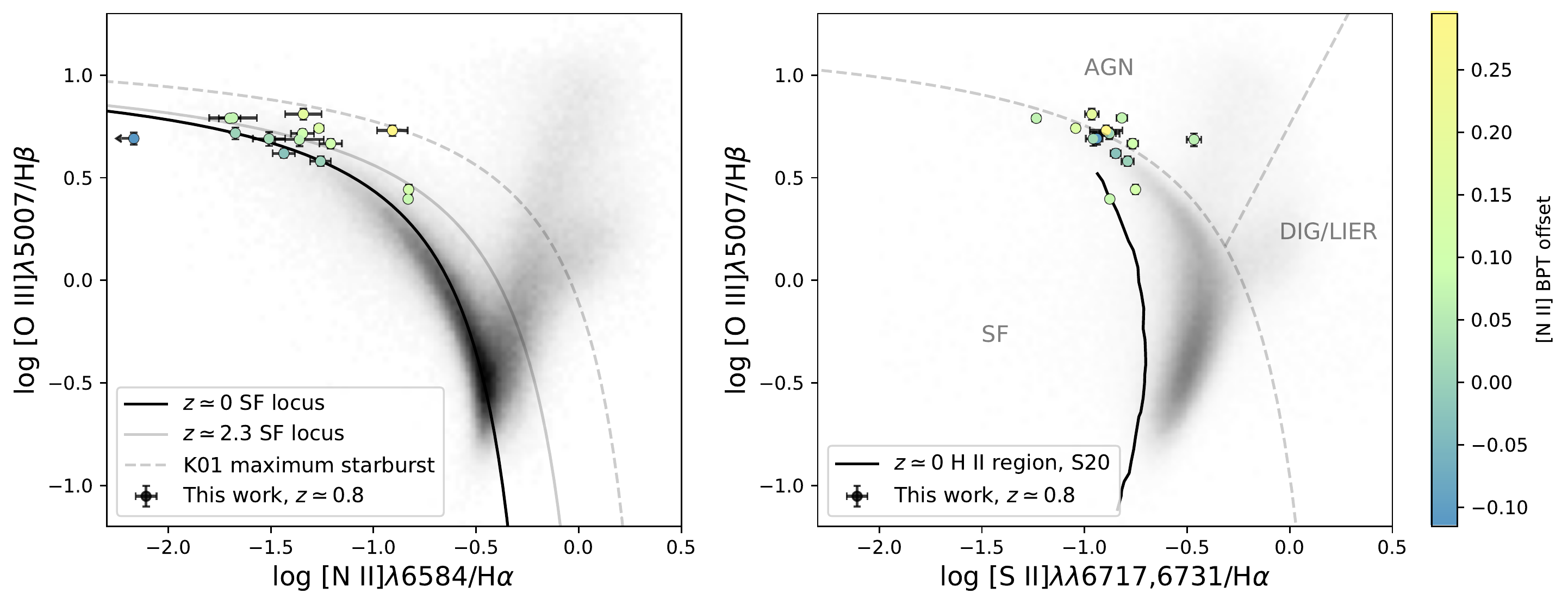}
  \caption{\textit{Left}: \Nii-\Oiii\ BPT diagram for our $z\sim0.8$ sample, color coded by offset from the \cite{Kewley_2013a} $z\sim0$ star forming locus (solid black line), where negative offsets indicate the target falls below the star forming locus. Higher redshift galaxies are significantly offset from the $z\sim0$ star forming locus in the \Nii\ BPT diagram, as demonstrated by an equivalent locus at $z\simeq2.3$ from the MOSDEF Survey \citep[solid gray line:][]{Shapley_2015}. All of the HST targets lie below the maximum starburst line \cite[][K01]{Kewley_2001}. The grayscale 2D density histograms in both panels show the same sample of $z\sim0$ galaxies from SDSS \citep{SDSS_DR7}.
  \textit{Right}: \Sii-\Oiii\ BPT diagram for the HST sample. Gray dashed lines show the separation into SF, AGN, and DIG/LIER regions at $z\sim0$ following the demarcations of \cite{Kewley_2001, Kewley_2006}. The solid black line shows the median relation of local \Hii\ regions from \citet{Sanders_2020a} based on observations from the CHAOS survey \citep{Berg_2015,Croxall_2015,Croxall_2016}.
We note that the S20 line shows the median relation from pure \Hii\ regions (i.e., lacking any DIG emission) and is offset from the galaxy-integrated SDSS sample, which contains DIG contributions.
  The HST targets are also offset from the median $z=0$ \Hii\ region sequence in the \Sii-\Oiii\ BPT plane, falling along the typical locus of luminous compact star forming galaxies at $z\sim0$.
  }
  \label{fig:bpt}
\end{figure*}

\subsection{Sample}
\label{sec:sample}

An overview of our HST sample is presented in Table \ref{tab:target_sample}. 
Our targets are a subset of the ``\Te\ sample'' of 32 galaxies at $z\sim0.8$ from \cite{Jones_2015}, which were originally drawn from the DEEP2 Galaxy Redshift Survey (DEEP2; \citealt{Davis_2003}; \citealt{Newman_2013}). The \Te\ sample was selected to be suitable for direct \Te-based metallicity measurements, and unbiased with regard to detection significance of the key temperature-sensitive \Oiii$\lambda$4363 emission line. In brief, the selection criteria from \cite{Jones_2015} are: (1) coverage from at least \Oii$\lambda\lambda$3726,3729 to \Oiii$\lambda$4959 in the DEEP2 spectra; (2) no signs of nuclear activity in the DEEP2 spectra; (3) sensitive measurement of the line ratio R$_{\mathrm{[O~III]}}$ = \Oiii$\lambda$4363 / \Oiii$\lambda\lambda$4959,5007 with 1-$\sigma$ uncertainty $\leq$ 0.0025. The last requirement ensures that \Oiii$\lambda$4363 is detected at $\geq 3\sigma$ significance for the median R$_{\mathrm{[O~III]}}$ ratio of the sample. 
The sample selection results in a bias toward high emission line luminosity and high excitation (e.g., large \Oiii/\Hb\ ratios), relative to the star-forming populations at $z=0.8$ and $z\simeq0$ \citep[see discussion in Section~2.3 of][]{Jones_2015}. 
However, this selection is blind to the actual \Oiii$\lambda$4363 flux. Of the 32 galaxies in the parent \Te\ sample, \Oiii$\lambda$4363 is detected at a median significance of 5.3$\sigma$ (with 6 galaxies having $<$ 3$\sigma$ significance). The \Te\ sample galaxies have typical $\Mstar \simeq 10^{8.5} - 10^{9.5} ~\Msun$ and span a narrow range of redshift $z=0.72-0.86$.

\subsection{Keck near-IR spectroscopy}
\label{sec:keck_spectra}

We obtained followup near-infrared spectroscopy of 23 galaxies from the \Te\ sample in order to measure the \Ha, \Nii, and \Sii\ emission lines, using the Keck/NIRSPEC and Keck/MOSFIRE spectrographs (Jones et al. in prep), used to measure the BPT offsets of our sample. While weather conditions prevented us from securing near-IR spectra of all targets, the observed subset is representative of the \Te\ sample. 
Relative to the broader $z\simeq0.8$ star forming galaxy population, this sample has large emission line equivalent widths and specific SFRs, relatively blue continuum colors, and luminosities representative of the overall sample observed by the DEEP2 survey \citep{Jones_2015}.
In brief, data were reduced using the LONGSLIT\_REDUCE pipeline written by George Becker in the case of NIRSPEC, and the MOSFIRE Data Reduction Pipeline in the case of MOSFIRE. Both pipelines perform wavelength calibration, instrument signature removal, sky subtraction, and 1-D spectral extraction. MOSFIRE 1-D spectra were extracted using the {\sc bmep} program \citep{Freeman_2019} following the methods used in the MOSDEF survey \citep{Kriek_2015}. 
Standard star observations taken on the same nights were used for telluric correction. Emission line fluxes were measured using single Gaussian fits \citep[as in][]{Jones_2015}. 
Figure~\ref{fig:keck_obs} shows an example Keck/MOSFIRE spectrum and best-fit emission line profiles, with good spectral resolution of the \Nii\ and \Sii\ doublets. We selected the 15 galaxies with the largest \Sii\ emission fluxes for HST grism spectroscopy, which forms the basis of this study. Availability of moderate resolution ground-based spectra is an important aspect, as this provides the \Nii/\Ha\ ratio for our targets (which is not available from HST grism spectra alone). Although the \Sii\ flux selection may bias the HST subsample in terms of emission line ratios, we find that the 15 HST targets are representative of the parent sample in terms of location on the \Nii\ BPT diagram. Figure~\ref{fig:bpt} shows both the \Oiii/\Hb\ versus \Nii/\Ha\ and \Sii/\Ha\ BPT diagrams for the HST sample, color coded by their offset from the \Nii\ BPT diagram (hereafter referred to as ``BPT offset") with negative offsets lying below the typical $z\sim0$ star forming locus. Each target's BPT offset is calculated as the minimum distance from the $z\sim0$ star forming locus on the \Oiii/\Hb\ versus \Nii/\Ha\ BPT diagram, with a median offset of $\sim0.08$ dex. In both diagrams, our targets ($z\sim 0.8$) show a variety of offsets from where $z\sim0$ \Hii\ regions typically lie. This work explores the cause of the \Nii\ BPT offset by spatially mapping emission lines to distinguish between various potential mechanisms that may drive emission at high redshift.

\subsection{HST observations and data reduction}
\label{sec:HST_obs}\label{sec:datared}

\begin{figure}
  \includegraphics[width=\linewidth]{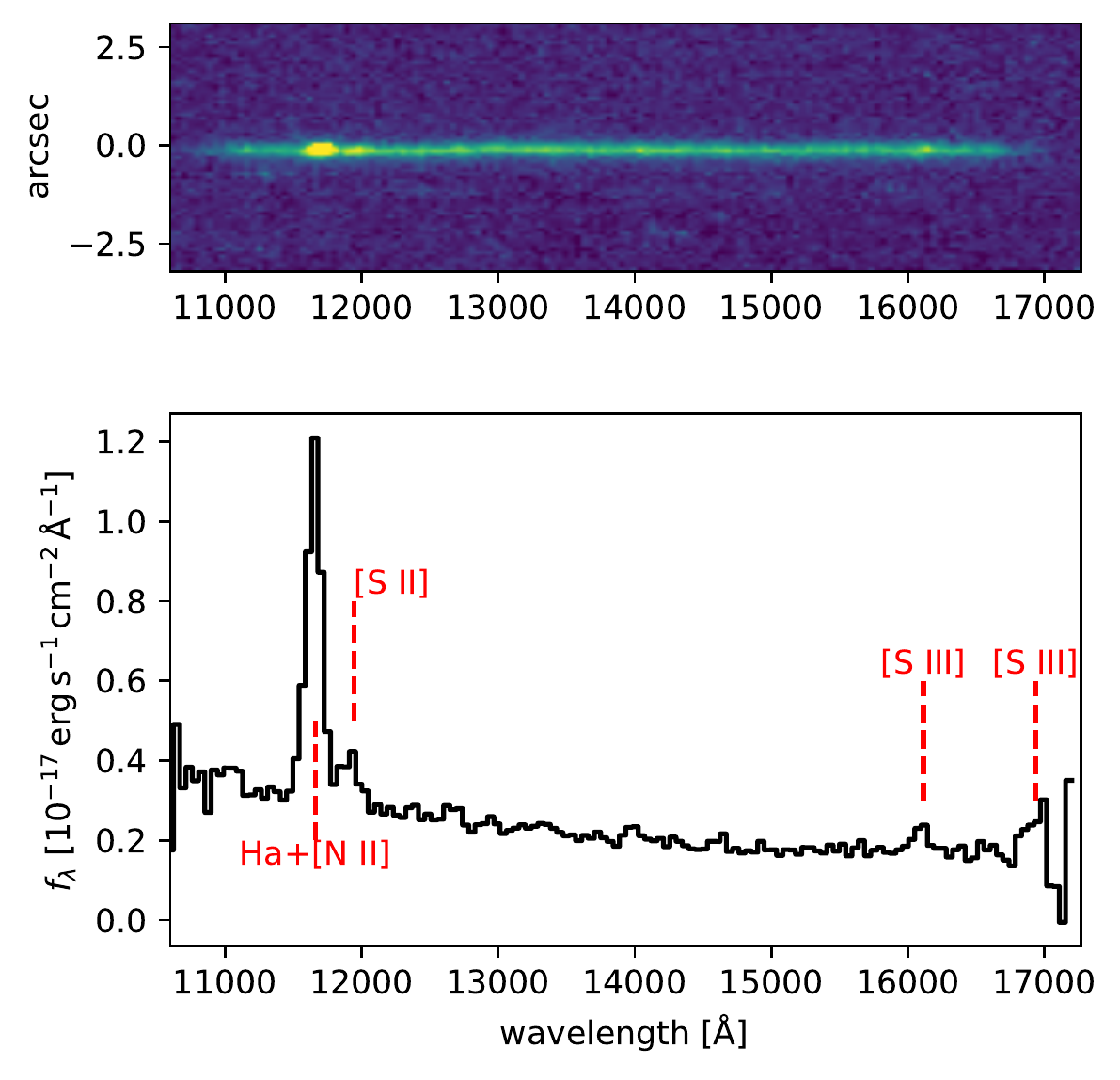}
  \caption{HST/WFC3-IR slitless grism spectrum of DEEP2-13043716 as processed by Grizli. \textit{Top}: 2D galaxy spectrum. \textit{Bottom}: Flux-calibrated 1D spectrum with our emission lines of interest highlighted: \Ha+\Nii, \Sii, and the \Siii\ doublet. Although \Ha+\Nii\ are blended in the HST grism spectra, as is the \Sii\ doublet, these lines are cleanly resolved in moderate resolution ground-based spectra available for all targets (e.g., Figure~\ref{fig:keck_obs}).} 
  \label{fig:2dspec}
\end{figure}

The targets were observed using HST's Wide Field Camera 3 (WFC3) via program GO-15077. Each target was observed for a single orbit, with 4 dithered G141 grism exposures plus 2 direct imaging exposures with the F140W filter, with a median exposure time of $\sim$2100 seconds per target. Photometric magnitudes are calculated from direct images and we assume a systematic error floor of 0.02 magnitudes (see Table \ref{tab:target_sample}).
We use the Grism Redshift and Line Analysis software (Grizli\footnote{https://github.com/gbrammer/grizli, version 0.8.0-4-g1153432}; \citealt{Brammer_2019}) to reduce the HST slitless grism spectroscopy data. Grizli offers a complete data reduction pipeline, beginning with preprocessing both the raw direct image and grism exposures, forward modeling their entire field of view, fitting redshifts through a spectral template synthesis, extracting both 1D and 2D spectra (e.g. Figure~\ref{fig:2dspec}), and outputting spatially resolved emission line maps and integrated emission line fluxes (given for our targets in Table~\ref{tab:EL_fluxes}). 
Of particular importance for this work is the level of contamination from overlapping spectra and zeroth-order images, which occurs with slitless grism data and can introduce errors in emission line measurements. In all cases the telescope orientation angle was constrained to avoid overlap of bright objects with our primary target spectra, and we additionally inspect the output spectra to check for contamination residuals. We do not find any cases where contamination significantly affects the primary target spectra which form the basis of this work. 

\begin{figure*}
  \includegraphics[width=\linewidth]{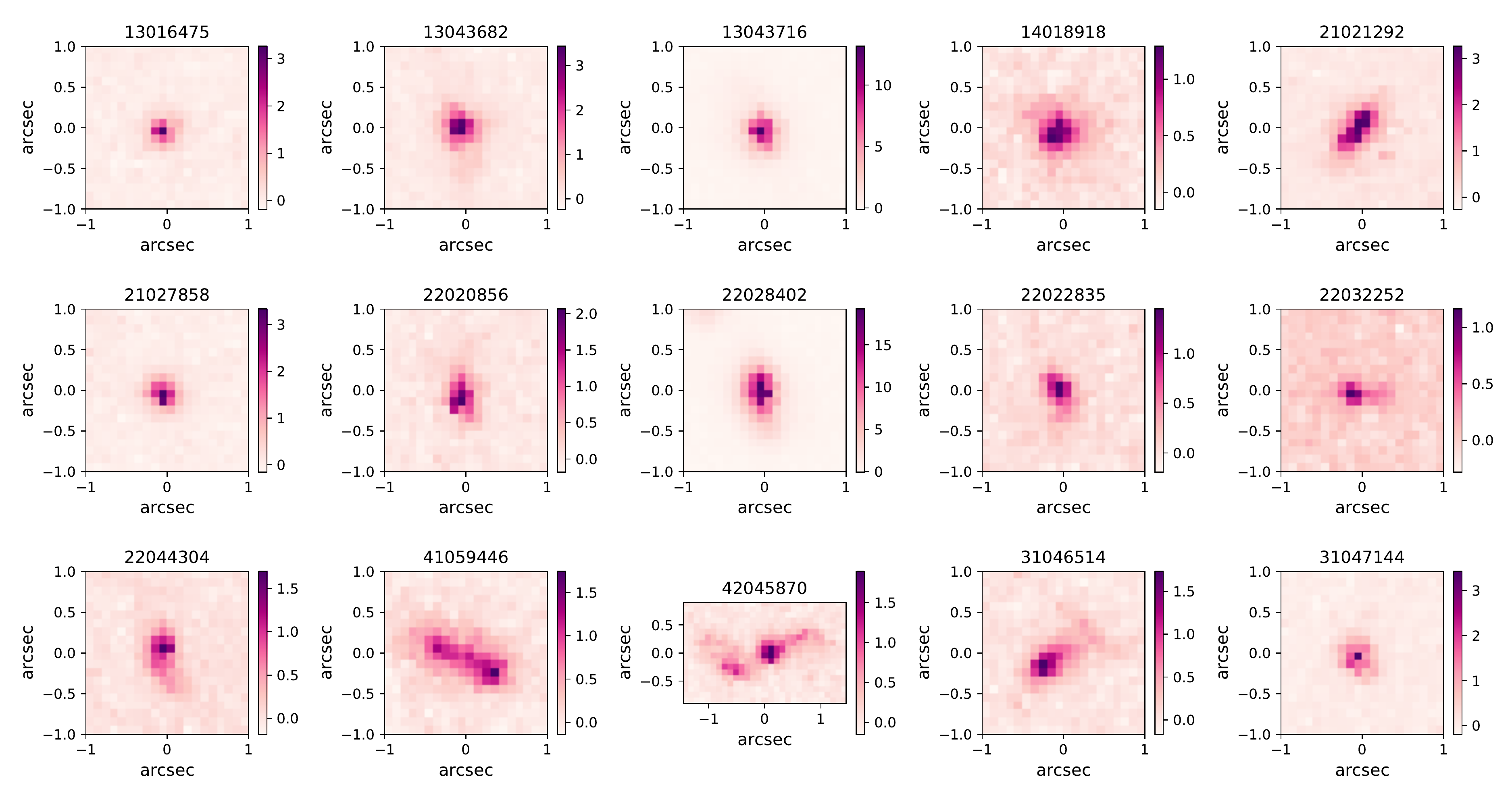}
  \caption{Direct HST F140W images for our HST sample. Nearly all targets have relatively compact morphologies, and even our most extended object, DEEP2-42045870, is free from \Ha+\Nii\ and \Sii\ blending effects. Colorbars have units of $10^{-20}~\mathrm{erg/s/cm^2/\AA}$.}
  \label{fig:direct_ps}
\end{figure*}

\begin{figure*}
  \includegraphics[width=\textwidth]{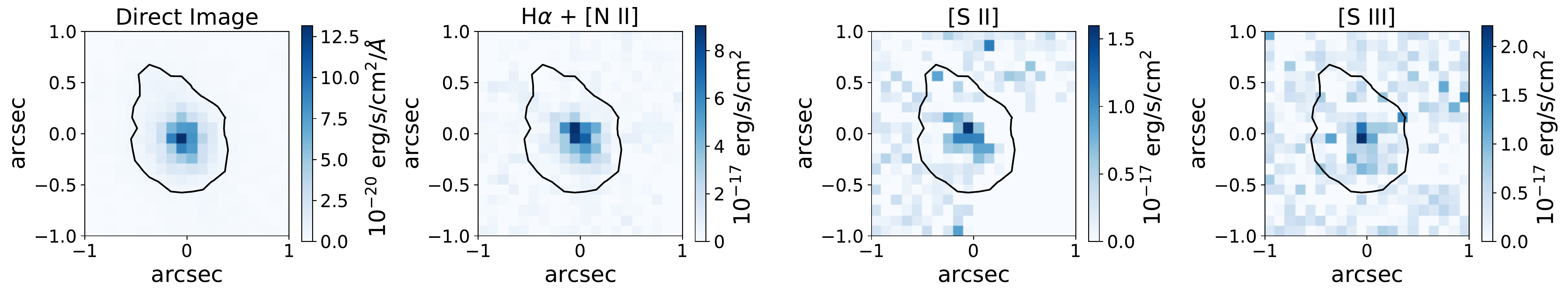}
  \caption{Raw Grizli emission line maps for DEEP2-13043716. From left to right: Direct image, \Ha+\Nii, \Sii, and \Siii\ emission line maps. The contours show the $\sim$ 25.5 magnitudes/arcsec$^2$ flux cutoff for defining galaxy extent.
  } 
  \label{fig:EL_maps}
\end{figure*}

In this paper we are mainly concerned with spatially resolved nebular emission line fluxes. Our analysis uses the \Ha+\Nii\ and \Sii\ emission line maps directly from Grizli, where both the \Ha+\Nii\ lines and the \Sii$\lambda\lambda$6717,6731 doublet are blended due to the grism's low spectral resolution. When both lines of the \Siii$\lambda\lambda$9068,9531 doublet are available, we use a weighted average of the doublet for our \Siii\ measurement, using the fixed ratio of 2.44:1 between the stronger ($\lambda$9531) and weaker ($\lambda$9068) lines.
When \Siii$\lambda$9531 is redshifted out of the G141 grism wavelength coverage and thus cannot be detected, we use only the \Siii$\lambda$9068 line multiplied by a factor of 2.44 (i.e. expressed as the expected intensity of the $\lambda$9531 line). Figures~\ref{fig:direct_ps} and \ref{fig:EL_maps} show examples of HST data products, showing direct images of the entire sample and spatially resolved emission line maps for one target (DEEP2-13043716), respectively. The pixel scale of the emission line maps is 0\farcs1, and the angular resolution is 0\farcs13 (full width at half-maximum). Emission line maps for the remainder of the sample can be found in Appendix~\ref{appendix}.

\section{Analysis} \label{sec:analysis}

This work utilizes the \Ha+\Nii, \Sii, and \Siii\ emission line maps output from Grizli to study spatially resolved emission line ratios. We are specifically interested in regions with elevated \SiiHa\ or \Siii/\Sii, which are indicative of potential effects of weak AGN, shocks and DIG, or escaping ionizing radiation as demonstrated in Section~\ref{sec:manga}. These effects can plausibly cause the observed deviation of high redshift galaxies from the $z\sim0$ star forming locus in BPT diagrams. 

We define the spatial extent of each target from its F140W direct image by applying a surface brightness cutoff for each system of $\sim$25.5 magnitudes/arcsec$^2$. This cutoff probes into the outer regions of our targets, including essentially all detected line emission while mitigating the amount of noise included from larger radii. 
This area corresponds to $\sim20-140$ independent spatial resolution elements per target (median of 40), indicating our galaxies are resolved and we have the appropriate resolution to distinguish spatial structure and trends with surface brightness. The targets are however sufficiently compact that there is no significant blending between the \Ha+\Nii\ and \Sii\ lines (see Figures \ref{fig:2dspec} and \ref{fig:direct_ps}), such that we obtain clean maps of both over the same spatial regions.

To increase the signal-to-noise ratio (SNR) for spatially resolved analysis, we utilize two methods of binning our data before calculating the line ratios of interest, namely \SiiHa\ and \Siii/\Sii. First we binned the data by \Ha+\Nii\ surface brightness in order to reach the lowest surface brightness regions that are not necessarily spatially contiguous. This is motivated by our objective to search for possible DIG or LIER emission (e.g., Figure \ref{fig:MANGA_dig_lier} shows prominent DIG at low surface brightness regions scattered throughout the galaxy). 
Second, to examine spatially correlated regions, we use the Voronoi binning method of \cite{Cappellari_2003}. Voronoi binning assigns spatially adjacent pixels to bins with a predetermined SNR requirement, and allows us to distinguish between the central and outer regions of galaxies more clearly. We required a median SNR threshold of 10 for \Ha+\Nii\ in each Voronoi bin.

\begin{figure*}
  \centering
  \includegraphics[width=\textwidth]{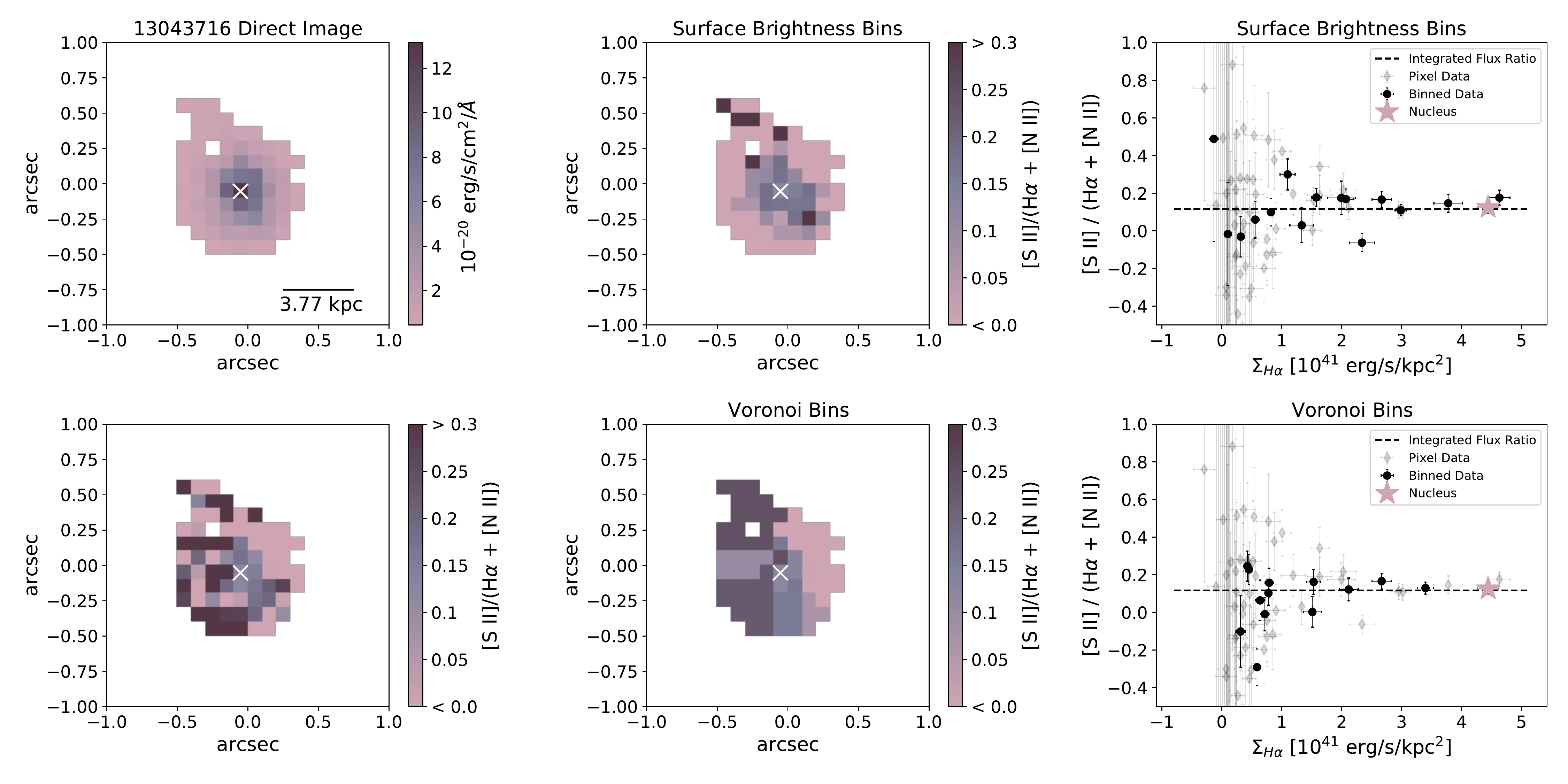}
  \caption{Spatially resolved \SiiHa\ observations for DEEP2-13043716. \textit{Top Left}: HST direct image (F140W). White crosses denote the nucleus, defined by the brightest central pixels in the direct image. \textit{Bottom Left}: The ratio of \SiiHa\ per pixel for this target. 
  \textit{Top Center}: \SiiHa\ map binned by \Ha+\Nii\ surface brightness. The pixels in this panel are color coded by the mean \SiiHa\ value in their \Ha+\Nii\ surface brightness bin; pixels in the same \Ha+\Nii\ surface bin are shown with the same \SiiHa\ ratio. \textit{Top Right}: Mean values of \SiiHa\ in each surface brightness bin plotted against \Ha+\Nii\ surface brightness.
  \textit{Bottom Center}: \SiiHa\ map binned into Voronoi bins. Our approach here is to display the same pixels as the original map, with each pixel color coded by the mean \SiiHa\ value in their respective Voronoi bin; pixels in the same Voronoi bin are shown with the same \SiiHa\ ratio. \textit{Bottom Right}: Mean values of \SiiHa\ in each Voronoi bin plotted against \Ha+\Nii\ surface brightness.
 Gray points in the right-hand panels show individual pixel values. The black data points show the ratio in each surface brightness or Voronoi bin. Pink stars denote the galaxy nucleus. Black dashed lines show the integrated \SiiHa\ ratio. Data points in the right-hand panels remain considerably flat with values \SiiHa~$\lesssim0.2$, ruling out LINER AGN, shocks, or DIG as significant contributors to the observed emission.} 
  \label{fig:analysis_bin}
\end{figure*}

\begin{figure*}
  \centering
  \includegraphics[width=\textwidth]{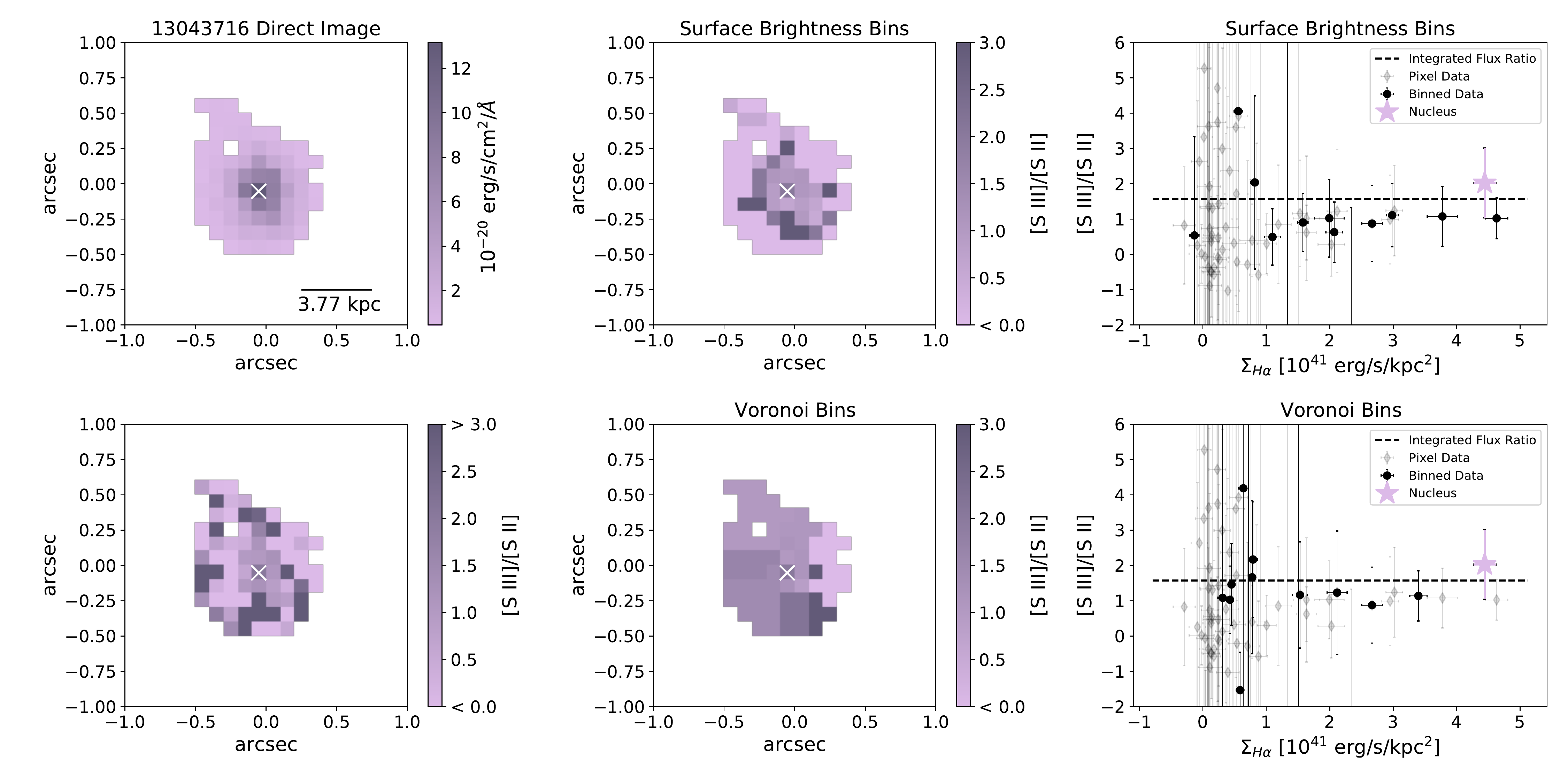}
  \caption{Spatially resolved \Siii/\Sii\ observations for DEEP2-13043716. \textit{Top Left}: HST direct image (F140W). White crosses denote the nucleus, defined by the brightest central pixels in the direct image. \textit{Bottom Left}: The ratio of \Siii/\Sii\ per pixel for this target. 
  \textit{Top Center}: \Siii/\Sii\ map binned by \Ha+\Nii\ surface brightness. The pixels in this panel are color coded by the mean \Siii/\Sii\ value in their \Ha+\Nii\ surface brightness bin; pixels in the same \Ha+\Nii\ surface bin are shown with the same \Siii/\Sii\ ratio. \textit{Top Right}: Mean values of \Siii/\Sii\ in each surface brightness bin plotted against \Ha+\Nii\ surface brightness.
  \textit{Bottom Middle}: \Siii/\Sii\ map binned into Voronoi bins. Our approach here is to display the same pixels as the original map, with each pixel color coded by the mean \Siii/\Sii\ value in their respective Voronoi bin; pixels in the same Voronoi bin are shown with the same \Siii/\Sii\ ratio. \textit{Bottom Right}: Mean values of \Siii/\Sii\ in each Voronoi bin plotted against \Ha+\Nii\ surface brightness.
 Gray out points in the right-hand panels show individual pixel values. The black data points show the ratio in each surface brightness or Voronoi bin. Purple stars denote the galaxy nucleus. Black dashed lines show the integrated \Siii/\Sii\ ratio. Data points in the right-hand panels remain considerably flat, and do not appear elevated in or near the brightest regions, ruling out density-bounded \Hii\ regions (with large ionizing radiation escape fractions) as a significant contributor to the observed emission.} 
  \label{fig:SIII_SII_analysis_bin}
\end{figure*}

Here we briefly discuss results for the galaxy DEEP2-13043716 demonstrated in Figures \ref{fig:analysis_bin} and \ref{fig:SIII_SII_analysis_bin}, as an example of what can be learned from the grism spectroscopy. This galaxy is offset by $\sim$0.1~dex from the $z\sim0$ star forming locus in the BPT diagram, slightly above the median offset in our sample (Table~\ref{tab:target_sample}). 
The ratio \SiiHa~$\simeq0.12$ is remarkably constant across an order of magnitude in surface brightness, thus ruling out LINER AGN, shocks and DIG as the dominant forms of emission on kpc scales. The \Siii/\Sii\ ratio for this object is broadly constant with values $\sim$1--2, although the error bars are large for regions of lower surface brightness. However in our regions of interest (i.e., those with the highest surface brightness), the \Siii/\Sii\ ratio remains flat and we can rule out a large contribution of density-bounded \Hii\ regions to the observed emission. The values are comparable to local high-excitation \Hii\ regions and $z\sim1.5$ main sequence galaxies \citep[e.g.,][]{Sanders_2020a}. 
In sum, this target lacks the clear distinguishing features in the line ratio maps which we would expect for LINER AGN, shocks, DIG, or escaping ionizing radiation, thus ruling out strong contributions from these sources. In contrast, this target exhibits relatively uniform \SiiHa\ and \Siii/\Sii\ ratios across all surface brightnesses which are consistent with that expected from excitation by newly-formed stars, suggesting that star formation may be the primary driver of its emission lines in all well-detected spatial regions. We perform a quantitative analysis for the complete sample in Section~\ref{sec:discussion}.

\section{Emission Line Powering Mechanisms}
\label{sec:discussion}

In this section we explore the possible sources of energy giving rise to the observed emission lines in our sample. 
Our approach here is to consider evidence for each emission mechanism based primarily on aggregate characteristics of the sample as a whole, as well as relevant individual objects (Appendix~\ref{appendix}). We note that the general results of this section also hold if we consider only the subset of galaxies with largest offsets in the BPT diagram (e.g., $>0.1$ dex offset from Table~\ref{tab:target_sample}).

\subsection{Weak AGN}

\subsubsection{Seyfert AGN}

\begin{figure}
  \centering
  \includegraphics[width=0.5\textwidth]{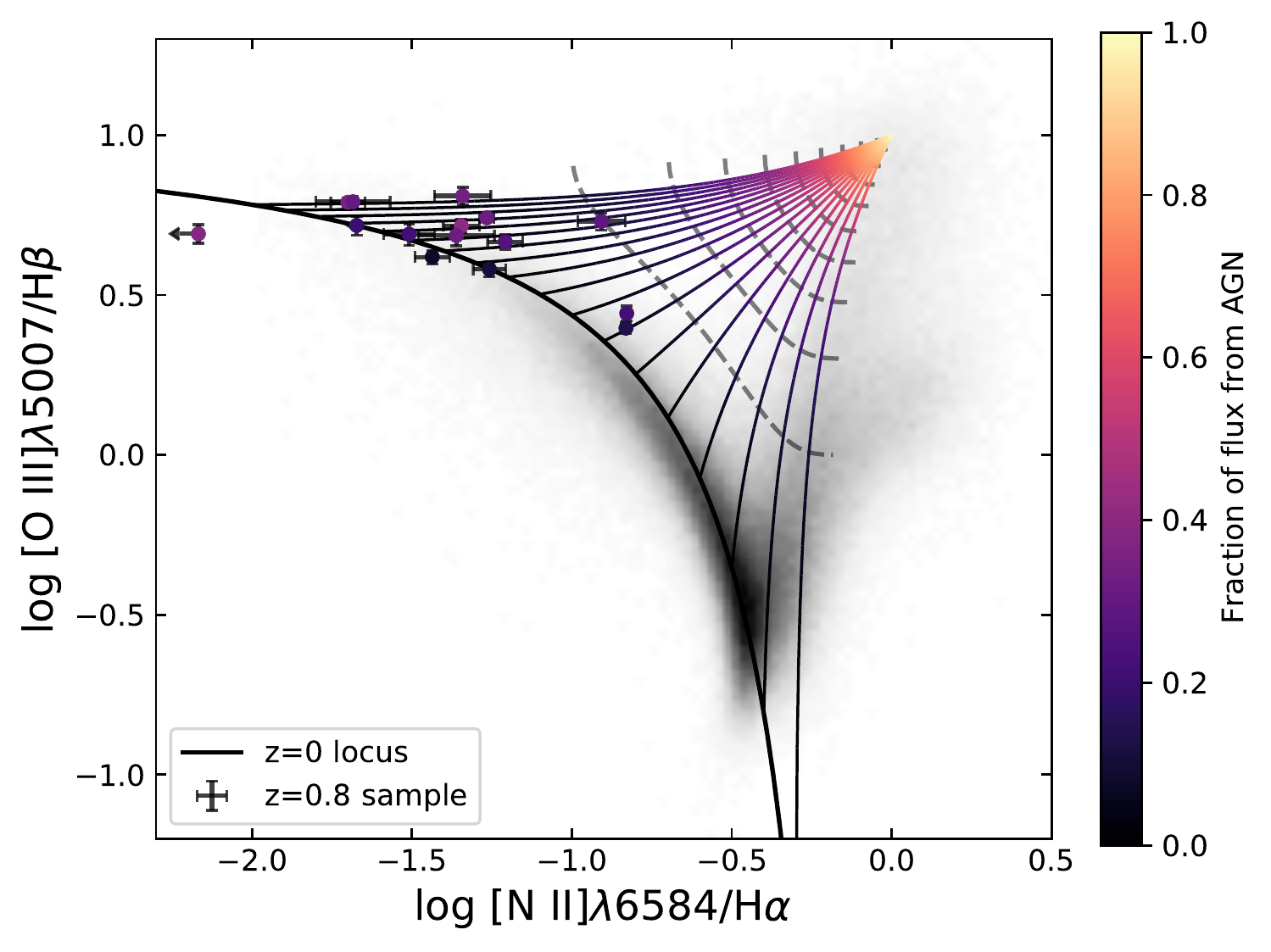}
  \caption{\Nii-BPT diagram showing how flux contribution from a fiducial Seyfert AGN offsets galaxies from the star forming locus. Colored lines show the fraction of mixing from purely star forming to pure AGN. Dashed contours show fixed mixing fractions starting 
  with 0\% AGN fraction at the star forming locus, increasing in steps of 10\% to 100\% AGN contribution to the \Ha\ flux. We overlay our HST targets, noting that the observed offsets can be explained by at most 10\% of the total flux coming from AGN. HST points are color coded by the percentage of \Ha\ flux in the central resolution element, ranging from 10-40\%. To explain the observed BPT offsets, $\sim$25\% of the \Ha\ flux in the central resolution element would have to be attributed to AGN emission.}
  \label{fig:SF_AGN_mixing}
\end{figure}

As noted in Section \ref{sec:manga}, we cannot distinguish Seyfert AGN by looking at \SiiHa\ levels because Seyfert AGN-dominated nuclei exhibit the same range of \SiiHa\ values as the spaxels dominated purely by star formation (see Figure \ref{fig:MANGA_seyfert}). However, we can still explore whether these offsets are driven by significant Seyfert AGN activity by estimating what percent of the flux must be attributed to AGN in order to cause the offsets observed in our HST sample. 

Figure \ref{fig:SF_AGN_mixing} shows how points along the \cite{Kewley_2013a} $z\sim0$ star forming locus would be offset when increasing fractions of the flux are from AGN, assuming 0\% Seyfert AGN contribution along the star forming locus and 100\% Seyfert AGN contribution at the tip of the AGN branch in the \Nii\ BPT diagram, assuming fiducial Seyfert line ratios of log(\Nii /\Ha)=0 and log(\Oiii/\Hb)=1 \citep[e.g.,][]{Kewley_2006}. Using this metric, BPT offsets in our sample can be explained by a relatively small contribution of $\lesssim$10\% of their \Ha\ flux coming from AGN. The percentages of \Ha\ flux in the central resolution elements of our sample range from 10--40\%. Therefore it is plausible that Seyfert AGN could be responsible for the observed offsets, if such emission represents a large fraction of flux from the central 1~kpc$^2$. If such AGN emission is common, we would expect to see signatures in IFS data that resolve the \Nii/\Ha\ lines, or in X-ray or radio observations. We note that, given the luminosities of our sample, we would need relatively deep X-ray or radio data to determine if there is significant AGN emission.
However, sensitive IFS surveys at similar redshift indicate low fractions of AGN activity at stellar masses of $\lesssim 10^{10}~\Msun$ and integrated \Nii/\Ha~$<0.2$ ratios of our sample \citep[with AGN signatures in $<$10\% of galaxies, e.g.,][]{ForsterSchreiber_2019}. Data with adaptive optics in particular can clearly distinguish elevated nuclear \Nii\ emission from AGN at the level required to explain offsets in our sample, but is rarely seen at these masses \citep[e.g.,][]{Jones_2013, Leethochawalit_2016, Hirtenstein_2019}. Thus we view Seyfert AGN activity as an unlikely explanation for the BPT offsets.

\subsubsection{LINER AGN}

\begin{figure*}
  \includegraphics[width=\textwidth]{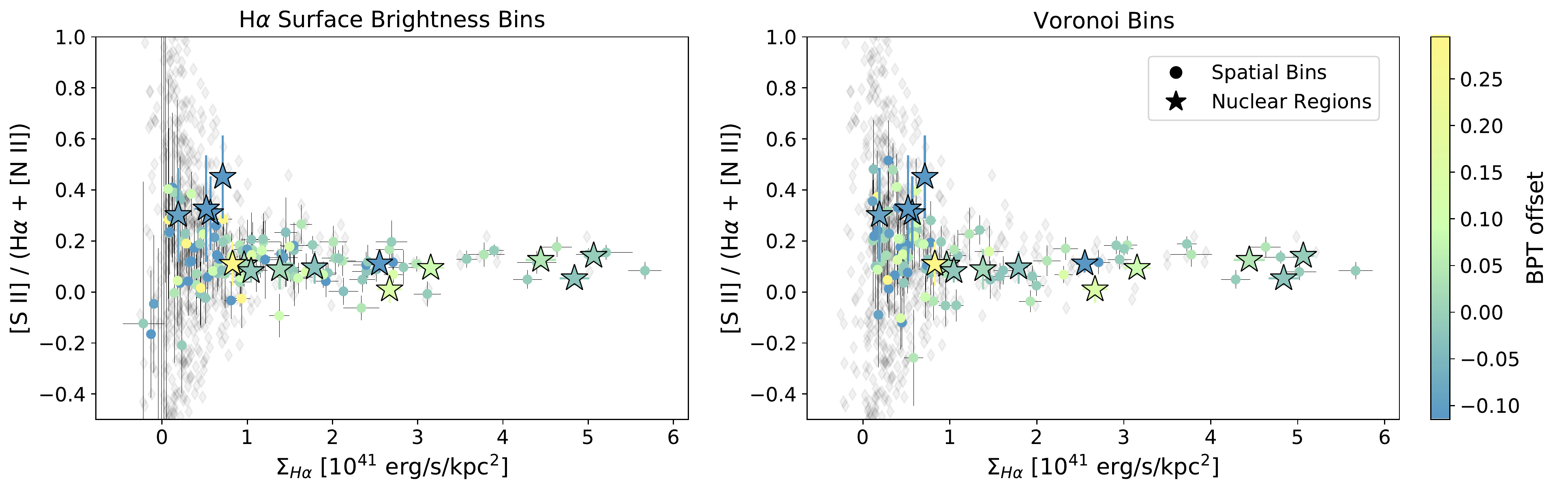}
  \caption{Our main diagnostic of spatially resolved \SiiHa\ vs. \Ha+\Nii\ surface brightness, for the entire sample plotted together (individual galaxy breakdowns can be found in Figure~\ref{fig:analysis_bin} and Appendix~\ref{appendix}). 
  Gray points represent individual pixel values, while colored circles represent spatial bins, and stars represent nuclear regions.
  Data points are color coded by the BPT offset of their respective galaxy. Both of our binning methods are represented here, binned by \Ha+\Nii\ surface brightness (left) and Voronoi bins (right). For both binning methods, the \SiiHa\ remains flat in the highest surface brightness regions, indicating no overall trend of contribution from AGN.
In comparison to the MaNGA galaxies shown in Figure~\ref{fig:MANGA}, although the HST sample is binned over larger regions (typical bin areas of $\sim$2 kpc$^{2}$), we are still able to distinguish between DIG and LINER emission on these scales in the MaNGA sample. The average \SiiHa\ ratios in the HST sample are also lower than the MaNGA galaxies (even when restricting to their \Hii\ regions), which we attribute to a combination of higher excitation and lower metallicity in these $z\simeq0.8$ HST targets.
  } 
  \label{fig:summary}
\end{figure*}

We can distinguish contributions from LINER type emission by looking for elevated \SiiHa\ in the nuclear regions (as shown in Figure \ref{fig:MANGA_liner}). Figure~\ref{fig:summary} shows all data for the aggregate sample, where we can see that the nuclear and highest surface brightness regions are relatively flat with values \SiiHa~$\lesssim0.2$, well below that expected for LINER emission. 
In order to explain the BPT offsets in our sample, LINER emission would have to account for $\sim$25\% of \Ha\ flux in the central resolution element as found for Seyfert AGN. For a fiducial mix of 75\% star formation (with \SiiHa~$=0.15$) and 25\% AGN (with \SiiHa~$=0.6$), we would expect a total \SiiHa~$\sim0.26$ in the nucleus. This is clearly inconsistent with the majority of our sample, thus ruling out LINER emission as a dominant source even within the central resolution elements of our targets.

Within Figure~\ref{fig:summary} there are four individual targets which show elevated \SiiHa~$\gtrsim0.3$ in their nuclear regions: DEEP2 IDs 21021292, 22022835, 41059446 and 42045870. Separate data for these galaxies is presented in Appendix~\ref{appendix} along with the rest of the sample. In three cases the elevated line ratios are not statistically significant ($\lesssim 1 \sigma$) suggesting that the ratios are likely due to noise fluctuations rather than AGN. 
While these nuclear line ratios are indicative of LINER AGN activity, all four galaxies in question have negligible offsets from the $z\simeq0$ BPT star forming locus: -0.001, 0.001, 0.016, and -0.03 dex, respectively (all consistent with zero offset).

We conclude that there is little evidence for LINER emission in the grism spectra, with only one object plausibly showing the expected LINER signature at $>1\sigma$ significance. In particular there is no sign of LINER activity among the galaxies exhibiting substantial offsets in the BPT diagram, despite sufficient angular resolution and sensitivity to detect the expected signal. Therefore, we can confidently rule out LINER-like line emission as the cause of BPT offsets in our sample.

\subsection{Shocks and Diffuse Ionized Gas}
\label{sec:discussion_DIG}

To determine if shocks or DIG may contribute significantly to the emission properties of our sample, we next look for evidence of elevated \SiiHa\ outside of the most luminous star forming regions in galaxy outskirts and in regions of low \Ha+\Nii\ surface brightness.
Although shocked gas and DIG emission show similar signatures in our diagnostic, they have different effects on different emission lines and therefore on BPT diagram offsets. Shocks can cause an elevated \Nii/\Ha\ ratio, for example as observed in a lensed $z\simeq1$ arc by \cite{Yuan_2012}. DIG-dominated regions are not likely to cause offsets on the \Nii\ BPT diagram, but have a significant effect on the \Sii\ flux and corresponding BPT diagram \citep[e.g.,][]{Sanders_2017}. We note that while DIG may not be responsible for causing \Nii\ BPT offsets, we still seek to quantify the role of DIG emission at moderate redshifts in our spatially resolved sample. 

Looking at the entire sample in Appendix~\ref{appendix}, the \SiiHa\ ratio appears flat in the lowest \Ha+\Nii\ surface brightness regions, to within the statistical uncertainties. While there are a few stray bins with higher \SiiHa\ ($\sim$ one bin per galaxy), the uncertainties with these data points are much higher and thus do not suggest significant contributions from shocks or DIG. The integrated data agree with the spatially resolved bins, showing offsets along the $z\sim0$ SF/AGN dividing line, with no clear signs of offset towards typical shock/DIG emission regions (see Figure \ref{fig:bpt}). While there is one galaxy more substantially offset towards the DIG/LIER region in \Sii/\Ha, its \Nii/\Ha\ ratio does not show a significant BPT offset.

\begin{figure}
  \includegraphics[width=0.5\textwidth]{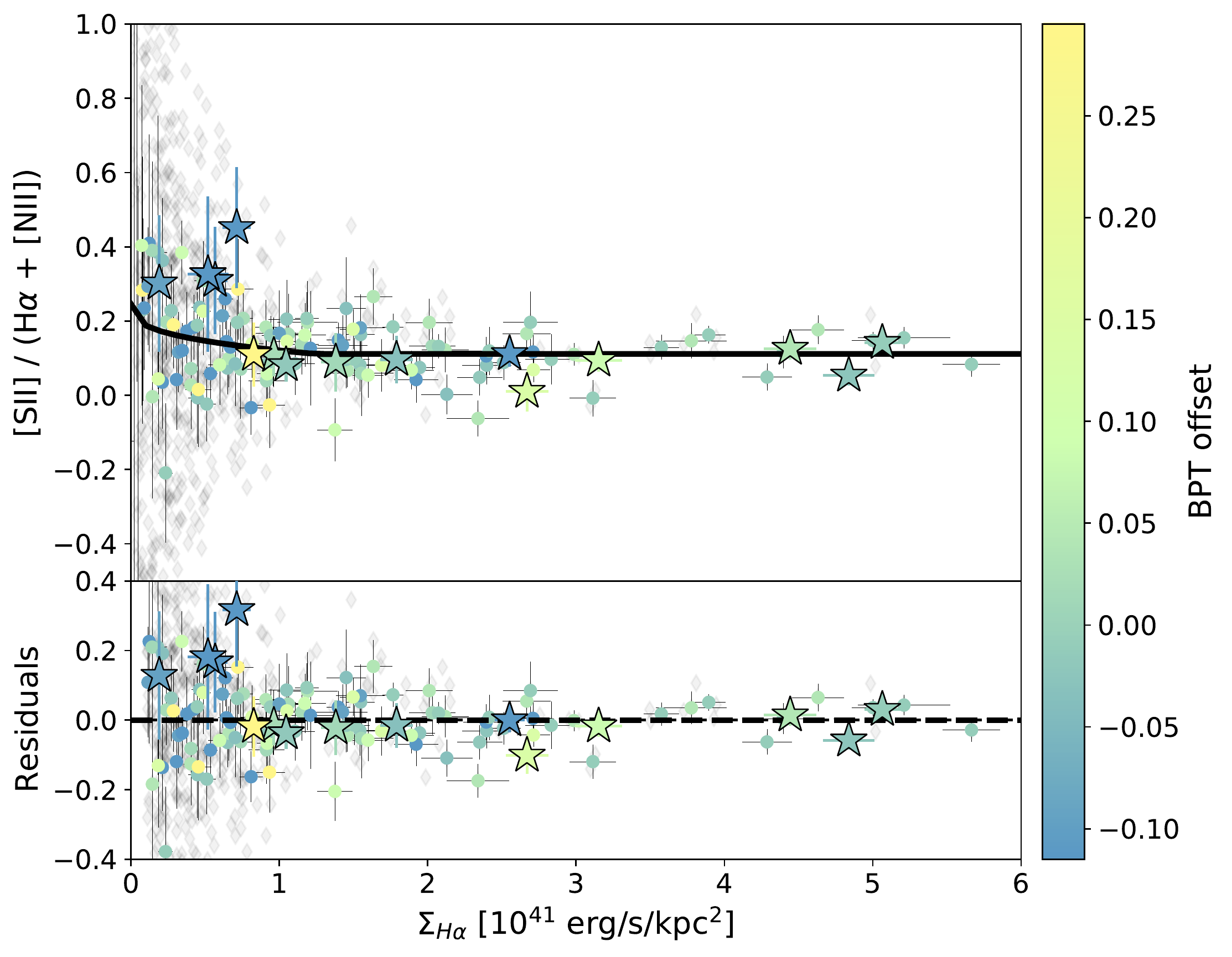}
  \caption{Our HST sample in aggregate (binned by \Ha+\Nii\ surface brightness) with the black line showing best-fit \SiiHa, using DIG fractions from \cite{Oey_2007} (top panel). The higher \SiiHa\ at low \SigHa\ is due to contributions from DIG emission. The bottom panel shows residuals from this best-fit relation. We estimate the total DIG fraction in our sample to be $\sim22\%$. The effect of bulk DIG emission at low surface brightnesses is marginally significant at the $2\sigma$ level. This DIG fraction, while consistent with measured DIG fractions at these surface brightnesses, is not large enough to cause the BPT offsets observed in our sample.}
  \label{fig:summary_DIG}
\end{figure}

While we do not detect strong DIG or shocked gas emission in the individual galaxies, we look to the combined sample for bulk DIG emission across all targets. In Figure~\ref{fig:summary_DIG} we fit our observed aggregate sample \SiiHa\ as a function of surface brightness binning by \Ha+\Nii\ surface brightness for increased SNR at the lowest surface brightnesses) with DIG fractions from \cite{Oey_2007}. 
We assume a functional form
\begin{equation}
    f_{DIG} = -1.50\times 10^{-14} \times \Sigma_{H\alpha}^{1/3} + 0.748
\end{equation}
as found by \cite{Sanders_2017}, treating the line ratios in SF and DIG regions as free parameters. We expect higher \SiiHa\ at lower surface brightnesses due to increasing DIG fraction.
Although the \cite{Sanders_2017} fit parametrizes $f_{DIG}$ as a function of \Ha\ surface brightness, using the blended \Ha+\Nii\ in this analysis should not strongly affect our results. We have verified that accounting for the contribution of \Nii\ surface brightness, with a plausible range of \Nii/\Ha\ ratios in DIG-dominated and star forming regions, yields results consistent within our 1$\sigma$ uncertainties.

Indeed, with this formalism we find evidence for the bulk effect of weak DIG emission at $\sim2\sigma$ significance (Figure~\ref{fig:summary_DIG}), with best-fit parameters of \SiiHa$=0.29 \pm 0.09$ for pure DIG emission and \SiiHa$=0.11 \pm 0.01$ for pure star forming regions.
Using these best-fit values with the total \Sii\ and \Ha+\Nii\ fluxes in all bins, we estimate the DIG fraction for our sample as a whole to be $\sim22\%$.
We note that the best-fit DIG line ratio of \SiiHa\ for a pure DIG region is somewhat lower than expected based on nearby galaxies, although compatible given the uncertainty \citep{Belfiore_2016, Zhang_2016, Sanders_2017}. If the true DIG line ratio is higher than given by this fit, then the DIG fraction of the sample would be lower than implied above (e.g., a typical DIG line ratio of 0.4--0.5 would imply only $f_{DIG}\sim10\%$ for the sample). 
Considering this wide range of plausible DIG line ratios values, our estimated DIG fractions remain consistent with \cite{Sanders_2017}, who suggest that DIG contributes $\sim$0-20$\%$ of the total flux at these surface brightnesses.
We conclude that while we do marginally detect weak DIG (at 2$\sigma$ significance) or shocked gas emission in our sample, it is not a dominant source of emission in these targets and is not responsible for their offsets in the BPT diagram.

\subsection{Escaping Ionizing Radiation}

Galaxies showing signs of escaping ionizing radiation are expected to exhibit elevated levels of \Siii/\Sii\ in or near the highest surface brightness regions \citep[e.g.,][]{Zastrow_2013,Alexandroff_2015}. 
Although our sample is not ideal for mapping \Siii\ (with the stronger $\lambda$9531 line redshifted beyond the G141 wavelength coverage in most cases), we consider the target with the strongest \Siii\ signal in Figure~\ref{fig:SIII_SII_analysis_bin}. As discussed in Section~\ref{sec:analysis} there is no significant evidence of elevated \Siii/\Sii\ in the regions of interest. We do not have sufficient \Siii\ detection for the remainder of the sample to comment further on whether escaping ionizing radiation may significantly affect the emergent nebular emission spectrum. However, the modest flux ratios of \Oiii/\Oii~$\lesssim5$ \citep{Jones_2015} suggest low ionizing escape fractions \cite[e.g., $f_{esc} < 0.1$ based on the results of][]{Izotov_2021}. 

While this study does not provide strong conclusions on the escape of ionizing radiation, we note that similar resolved mapping is possible at lower redshifts with HST's grisms, and with future facilities including the James Webb Space Telescope (JWST) and Nancy Grace Roman Space Telescope. Such emission line maps may be a powerful diagnostic, especially given the challenges of confirming robust Lyman continuum emission with deep high-resolution imaging \citep[e.g.,][]{Mostardi_2015}.

\subsection{Star Formation}

The spatially resolved line ratios in our sample, in addition to properties including morphology, color, and spectra (\citealt{Jones_2015}) indicate that spatially extended star formation plays a large and likely dominant role in the nebular emission properties. While we have largely ruled out significant contributions from AGN, shocks and DIG, the HST grism spectra are in excellent agreement with excitation by star formation. The spatially resolved line ratios are broadly indicative of star forming \Hii\ regions with moderate metallicity (with median \oh\ $=8.2$ as indicated by direct measurements using the \Oiiit\ feature; \citealt{Jones_2015}). The large surface brightness indicates a high star formation rate surface density (of order 1 M$_{\odot}$/yr/kpc$^2$) which is typical of star-forming galaxies at $z>1$. At such high surface brightness, DIG is expected to contribute minimally to the total line fluxes ($\lesssim$20\%; \citealt{Oey_2007,Sanders_2017, Shapley_2019}), consistent with the weak DIG emission signatures from Section \ref{sec:discussion_DIG}.

This conclusion remains consistent with other resolved line ratio work targeting \Hii\ regions at high redshift. For example,  \cite{Jones_2013} and \cite{Leethochawalit_2016} conclude that emission lines originate predominantly from star forming \Hii\ regions in their lensed samples of galaxies at $z\sim2$, and that their observed offsets are broadly consistent with extensions of the star-forming locus seen in other high redshift surveys \citep[e.g.,][]{Steidel_2014, Shapley_2015}. Similarly, \cite {Genzel_2014} and \cite{ForsterSchreiber_2019} find AGN to be uncommon in galaxies at similar masses at $z\simeq1-2$, namely less than 10\% for galaxies with $\log (\Mstar/\Msun) < 10.3$. 
Furthermore, star formation rates derived from \Ha\ and \Hb\ emission lines are in good agreement with results from other methods in $z>1$ galaxies \citep{Shivaei_2016}, supporting star formation as the dominant source of nebular line flux. Our spatially resolved analysis both strengthens the evidence that our sample is dominated by star formation, and provides stringent limits on the fraction of line flux which can be attributed to other ionization mechanisms.

\section{Emission line excitation from star forming regions at high redshift}
\label{sec:SF}

Analysis of our sample described in Section~\ref{sec:discussion} strongly suggests that the dominant source of line emission in all cases is from \Hii\ regions powered by star formation. Here we discuss the implied star formation properties of our sample and consider which physical properties of \Hii\ regions might be responsible for offsets in the BPT diagram compared to $z\sim0$ galaxies. 

The surface brightness of our sample spans $\Sigma_{H\alpha} \sim 0.5$--$5 \times 10^{41}$~erg\,s$^{-1}$\,kpc$^{-2}$ for well-detected regions, with total luminosities $L_{H\alpha} \sim 10^{42}$~erg\,s$^{-1}$ (uncorrected for extinction; two objects have significantly larger luminosities). The emission is extended over several kpc in all cases with at most $\sim$40\% of the flux being found within a single resolution element.
Accounting for $\sim$1--2 magnitudes of extinction (based on Balmer lines; \citealt{Jones_2015}), this corresponds roughly to SFR densities $\Sigma_{\mathrm{SFR}} \sim 1-10~\Msunyr$\,kpc$^{-2}$ and total SFR~$\sim10$--$50~\Msunyr$ (assuming a \citealt{Chabrier_2003} initial mass function). These values are characteristic of moderately massive ($\Mstar \sim  10^{9}~\Msun$) star forming galaxies at $z>1$ \citep{Speagle_2014, Shivaei_2015}.

Line ratios and positions in the BPT diagram are known to vary systematically with total emission line luminosity and surface brightness. At the relatively low luminosities (or equivalently SFRs) typical of spatially resolved $z\sim0$ spiral galaxies, lower $\Sigma_{H\alpha}$ correlates with offsets toward higher \Nii/\Ha, \Sii/\Ha, and \Oiii/\Hb\ which can be explained by increasing fractions of DIG (e.g., \citealt{Oey_2007, Zhang_2016, Masters_2016, Sanders_2017, Shapley_2019}). In contrast, the global trend is opposite, where {\em higher} integrated $L_{H\alpha}$ correlates with offsets toward higher \Nii/\Ha\ and/or \Oiii/\Hb\ as seen in the BPT diagram for both $z\sim0$ and $z>1$ galaxies \citep[e.g.,][]{Brinchmann_2008, Cowie_2016, Masters_2016}. Our sample is clearly characteristic of the latter case, having luminosities typical of the $z>1$ galaxies analyzed by \cite{Cowie_2016}. 
While our sample is not large enough to establish an explicit relationship in BPT offset with luminosity, we can nonetheless explore whether the observed BPT offsets of the HST targets follow the same trends as found in larger populations. The luminosity of our sample (typical $L_{H\beta} \sim 10^{41.5}$~erg\,s$^{-1}$) corresponds to offsets in \Nii/\Ha\ and \Oiii/\Hb\ of $\sim$0.1-0.2 dex according to \cite{Cowie_2016}. This is in good agreement with the median offsets solely in \Nii/\Ha\ or \Oiii/\Hb\ (0.19 and 0.7 dex, respectively) in our sample. Thus the BPT offsets in  Table~\ref{tab:target_sample} are commensurate with trends in global properties seen at both low and high redshifts. As such, the physical mechanisms responsible for offsets in this sample are likely to be the same as in broader galaxy populations.

Many other works have further explored the main driving mechanisms of the more extreme ISM conditions at $z>0$. Such studies have invoked a variety of properties at high redshift such as \Hii\ regions having higher ionization parameters than at $z\sim0$ (e.g., \citealt{Kewley_2013a, Kewley_2013b}), an enhanced N/O abundance (e.g., \citealt{Masters_2014, Shapley_2015}), and high $\alpha$/Fe abundance ratios (e.g., \citealt{Steidel_2016, Shapley_2019, Sanders_2020b, Sanders_2020c}). 
A key result of our spatially resolved analysis is that the emission line ratios appear nearly constant across individual galaxies. 
That is, offsets in the BPT diagram appear linked to physical conditions in the \Hii\ regions. Given the limited number of emission lines in HST grism data, we cannot clearly distinguish between various mechanisms proposed in the literature on spatially resolved scales. However, the results of this work support treating the integrated emission line fluxes of high-z galaxies as originating from \Hii\ regions ionized by massive stars, for purposes of modeling and interpreting their spectra. 
As an example, the photoionization modeling framework from \cite{Sanders_2020c} -- adopting the \Te-based O/H measurements for our sample \citep{Jones_2015} -- indicates that the BPT offsets in our sample are caused by super-solar O/Fe abundances.  In our future work, we will apply a rigorous photoionization modeling analysis of this sample using stellar population synthesis models as input (e.g. BPASS).

\begin{figure}
  \includegraphics[width=0.45\textwidth]{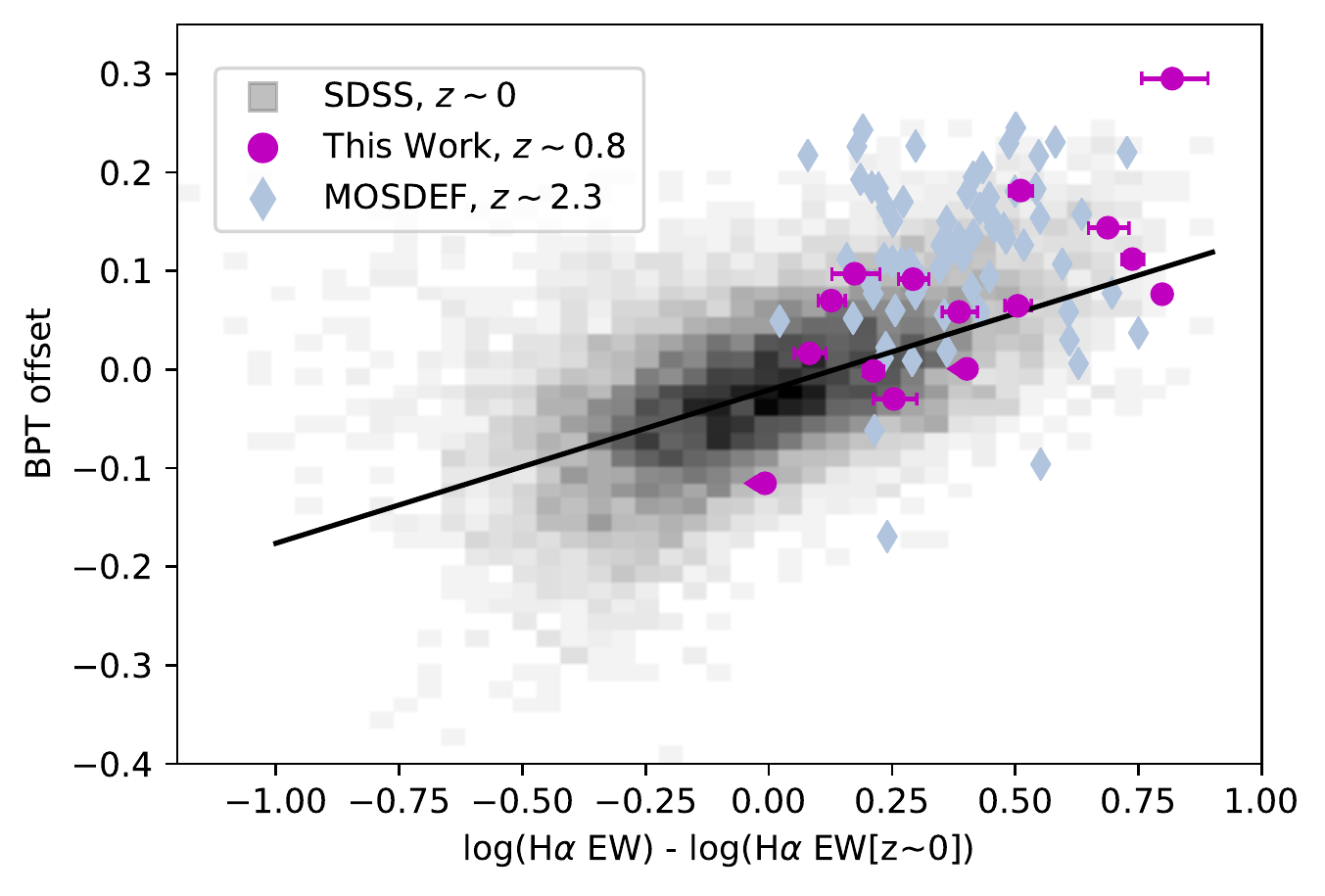}
  \caption{BPT offset plotted against normalized \Ha\ equivalent width for star forming galaxies at $z\sim0$ (SDSS), our HST sample at $z\sim0.8$, and $z\sim2.3$ (MOSDEF, \citealt{Reddy_2018, Sanders_2018}). The solid black line shows the best fit to the SDSS data. We see a clear trend in the SDSS data of BPT offset with normalized \Ha\ equivalent width. 
  The high redshift sources are, on average, offset above the best fit line at fixed equivalent width, consistent with expectations for increased $\alpha$-enhancement at earlier cosmic times. \Ha\ equivalent width is chosen here as a proxy for sSFR, which correlates with age. This is interesting to consider in the context of chemical abundance patterns, in particular super-solar $\alpha$/Fe at young ages. The correlation seen here indicates that larger BPT offsets are indeed associated with younger ages (higher sSFR and equivalent width), consistent with the trend expected if $\alpha$-enhancement is responsible for the observed BPT offsets in our sample.
  }
  \label{fig:HaEW_vs_bptoffset}
\end{figure}

Increased O/Fe abundances are generally expected for galaxies at high redshifts, given the shorter enrichment timescale for $\alpha$-elements such as O compared to Fe. Recent work has suggested that O/Fe may typically reach $\sim 4\times$ the solar value at $z\sim2$ \citep[e.g.,][]{Steidel_2016,Jones_2018,Sanders_2020c,Topping_2020b}. Such $\alpha$-enhancement results in harder stellar ionizing spectra at fixed O/H abundance, which drives offsets in the BPT diagram. While we cannot measure O/Fe directly in our sample, we expect it to correlate with young ages and hence high specific star formation rates (sSFR).  Figure~\ref{fig:HaEW_vs_bptoffset} explores this idea, using \Ha\ equivalent width (EW) as a proxy for sSFR (and hence possibly $\alpha$-enhancement). To correct for the dependence of \Ha\ EW along the $z\sim0$ BPT locus, we first fit for a relationship between mean \Ha\ EW of the $z\sim0$ sample and the parameter O3N2 ($= \log$(\Oiii/Hb)$-\log$(\Nii/\Ha)), using the rest frame \Ha\ EW. 
We note that lines of constant O3N2 are roughly orthogonal to the BPT star-forming sequence in the region of interest here, such that normalization at fixed O3N2 is appropriate for comparison with our BPT offset measurements. We find a linear relationship of
\begin{equation}
    \log(H\alpha~\mathrm{EW}[z\sim0]) = 0.762 + 0.779\times \mathrm{O3N2}
    \label{eq:EW_norm}
\end{equation}
(with \Ha\ EW measured in \AA) to be a good fit for star-forming galaxies in SDSS, spanning the O3N2 $>1.2$ values of our sample. 

Figure~\ref{fig:HaEW_vs_bptoffset} plots the normalized \Ha\ EW (using the mean relationship in Equation \ref{eq:EW_norm}) at fixed O3N2, revealing a clear correlation between normalized \Ha\ EW and offsets in the BPT diagram for the SDSS sample \citep[as has been noted previously by, e.g.,][]{Brinchmann_2008}. Moreover, the higher redshift data -- shown for both our HST sample at $z\sim0.8$ and MOSDEF galaxies at $z\sim2.3$ \citep{Reddy_2018, Sanders_2018} -- show increased BPT offsets at fixed EW. The median displacement above the SDSS best fit line appears to increase with redshift (0.02 dex for the HST sample, 0.08 dex for the MOSDEF sample), although there is no clear correlation within the MOSDEF sample. This suggests a smooth evolution of increased BPT offset with increasing redshift at fixed \Ha\ EW beyond the relation at $z\sim0$
(consistent with the expectation of increasing $\alpha$ enhancement at high redshifts compared to $z\sim0$).

While our analysis supports $\alpha$-enhanced abundance patterns, we note that this is indistinguishable from the case where ISM metallicity is higher than that of ionizing stars, with solar abundance patterns. Recent ISM enrichment from massive stars may show significant deviations from the mean galaxy metallicity and be localized to $\lesssim1$ kpc spatial scales \citep{Krumholz_2018}. 
While the stars responsible for ionizing \Hii\ regions are generally thought to have similar composition as the ISM, our data do not provide direct measurements of the stellar abundance.
We also note that \cite{Topping_2020b} suggest $\alpha$-enhancement even among $z\sim2$ galaxies which are not offset from the $z\sim0$ BPT locus, such that additional properties beyond abundance patterns must play a significant role.

We note that there are many other possibilities that may drive BPT offsets beyond abundances. In particular, effects of binary stellar evolution (leading to stripped stars or high-mass X-ray binaries), turbulence \citep[e.g.,][]{Gray_2017}, and higher ionization parameters \citep[e.g.,][]{Jaskot_2019} may also drive high redshift galaxies off of the $z\sim0$ star forming locus.
Nonetheless, our results are generally consistent with offsets driven by moderate $\alpha$-enhancement, if indeed the \Ha\ EW correlates with young ages and $\alpha$-enhanced abundance patterns. We conclude that photo-ionization modeling of \Hii\ regions ionized by massive stars will be a reasonable tool for further interpreting these results since there are no significant signatures from other ionizing sources beyond star formation.

\section{Summary}
\label{sec:summary}

In this paper we present and analyze spatially resolved emission line maps of 15 galaxies at $z \sim 0.8$ in the DEEP2 fields, observed with HST grism spectroscopy. Our targets show moderate offsets from the $z\sim0$ locus of star forming galaxies on the \Oiii/\Hb\ versus \Nii/\Ha\ BPT diagram, comparable to those seen in typical $z\sim2$ galaxies. 
While line emission in the sample appears to be dominated by star formation, we analyzed \Ha+\Nii, \Sii\ and \Siii\ maps in order to distinguish signatures of AGN, shocks/DIG, or escaping ionizing radiation in our sample to determine whether these mechanisms may be responsible for the observed BPT diagram offsets. 
Our main results are as follows:

\begin{enumerate}

\item \textit{AGN}: We achieve $\sim$1 kpc spatial resolution to distinguish possible contributions from faint AGN. 
We find that AGN would need to account for a large fraction of emission ($\gtrsim$25\% of \Ha\ flux) within the central resolution element, in order to explain the observed magnitude of BPT offsets. 
In the case of LINER AGN, we would expect significantly elevated \SiiHa\ ratios in the nuclear regions, which are not observed. In particular the objects in our sample with the largest BPT offsets are inconsistent with significant emission from LINER AGN. For Seyfert AGN, the \SiiHa\ ratio is not necessarily different from that of \Hii\ regions, and so we cannot clearly distinguish their presence in the sample. However if Seyfert AGN emission is indeed widespread, it would be easily distinguishable in IFS surveys via high \Nii/\Ha\ ratios in galaxy nuclei, which are not commonly observed in galaxies with mass and metallicity comparable to this sample. We therefore conclude that substantial Seyfert AGN contributions are unlikely, and LINER AGN are clearly ruled out as causes of BPT offsets. 

\item \textit{Shocks and DIG}: We examine how \SiiHa\ varies spatially and as a function of surface brightness, in order to distinguish contributions from shocks and DIG. There is no clear evidence of shock/DIG emission in individual galaxies, however we find some evidence of bulk DIG emission at $2\sigma$ significance across the entire sample. We estimate the DIG fraction in our sample to be $\sim22$\%, in line with the expected fractions based on the resolved surface brightnesses of our sample, and thus conclude that DIG or shocked gas emission are not the primary emission sources in our targets.

\item \textit{Escaping ionizing radiation}: In principle the \Siii/\Sii\ ratio can identify locations of density-bounded \Hii\ regions with escaping ionizing radiation. We demonstrate for one object that high ionizing escape fractions are unlikely, although in general the \Siii\ signals are too weak to examine this possibility. Nonetheless future deep grism spectroscopy mapping the \Siii\ lines, for example with JWST, holds promise for further study. 

\item \textit{Star formation}: All targets in our sample exhibit spatially extended emission with high surface brightness and line ratios indicative of star forming \Hii\ regions. It is clear that star formation is responsible for the vast majority of emission line flux in our targets. Having found that shocks, DIG, and LINER emission cannot explain the observed BPT offsets, we conclude that these line ratio offsets are most likely caused by different physical characteristics of \Hii\ regions and/or young stars compared to typical star forming galaxies at $z\sim0$. 
We consider trends in BPT offset with global properties of $z\sim0$ star-forming galaxies, and find that the $z\sim0.8$ sample follows similar relations. In particular the BPT offsets are consistent with expectations based on Balmer line luminosities and equivalent widths. The BPT offsets thus appear to be connected to high SFR and sSFR which are typical of high-$z$ galaxies. Although not conclusive, the trend with sSFR supports recent work suggesting that $\alpha$-enhanced abundance patterns cause BPT offsets at high redshift.

\end{enumerate}

Overall, the spatially resolved data in our sample appear to be consistent with emission driven by star formation, despite their integrated line ratios being offset from the $z\sim0$ BPT star forming locus. This highlights an important outstanding question in galaxy evolution studies: how do the physical properties of \Hii\ regions at high redshifts differ from those observed locally? 
In future work we intend to perform a thorough photo-ionization modeling analysis of this sample with a full suite of emission lines to further probe this question and discern between the properties that may drive line ratio offsets, such as high ionization parameters, enhanced N/O abundance, or high $\alpha$/Fe.
A key result of this work is that, since emission is powered by star formation, models of photoionization by massive stars should be appropriate for interpreting the spectra of high-$z$ galaxies with similar BPT offsets. Likewise, direct-method abundance measurements \citep[e.g.,][]{Jones_2015,Sanders_2020b} and other nebular emission probes should indeed reflect the physical conditions of \Hii\ regions.

Looking forward, we can soon extend this type of analysis to a wider range of redshifts and a larger sample size. JWST will provide increased SNR along with excellent spatial resolution through grisms, IFU, or slit-stepping to distinguish between the properties of nuclear regions, high surface brightness regions, and galactic outskirts. This will allow further exploration of DIG emission and trends with surface brightness, and to use \Siii\ as a diagnostic of density-bounded \Hii\ regions with escaping ionizing radiation at high redshifts. Additionally, this analysis is not limited by spectra that cannot resolve \Ha\ and \Nii; as shown with the MaNGA data in Figure~\ref{fig:MANGA}, we can clearly distinguish different physical mechanisms using the \SiiHa\ metric without high spectral resolution. Space-based grism spectroscopy is thus a powerful technique which can additionally benefit from multiplexing. 
While we found no strong evidence of properties beyond star formation in our sample, a larger spatially resolved sample can provide more certainty for high redshift galaxies as a broader population. However if high redshift galaxy emission is truly dominated by star formation, integrated emission line measurements are sufficient to discern between the various proposed properties within \Hii\ regions. Further understanding of the origin of emission line excitation at high redshift will improve our ability to calibrate metallicity measurements between the low and high redshift universe, and propel our knowledge of galaxy evolution. 

\begin{acknowledgments}
We thank the anonymous referee for a constructive report which substantially improved the content and clarity of this manuscript. This work is based on observations with the NASA/ESA Hubble Space Telescope obtained by the Space Telescope Science Institute, which is operated by the Association of Universities for Research in Astronomy, Incorporated, under NASA contract NAS5-26555. Support for Program number GO-15077 was provided through a grant from the STScI under NASA contract NAS5-26555. Support for RLS was provided by NASA through the NASA Hubble Fellowship grant \#HST-HF2-51469 awarded by the Space Telescope Science Institute, which is operated by the Association of Universities for Research in Astronomy, Incorporated, under NASA contract NAS5-26555. CLM acknowledges support from the National Science Foundation under grant AST-1817125. MCC acknowledges support from the National Science Foundation under grants AST-1815475 and AST-1518257.
\end{acknowledgments}

%






\appendix

\section{Resolved emission line measurements for the full sample}
\label{appendix}

\begin{figure*}
  \centering
  \includegraphics[width=0.95\textwidth]{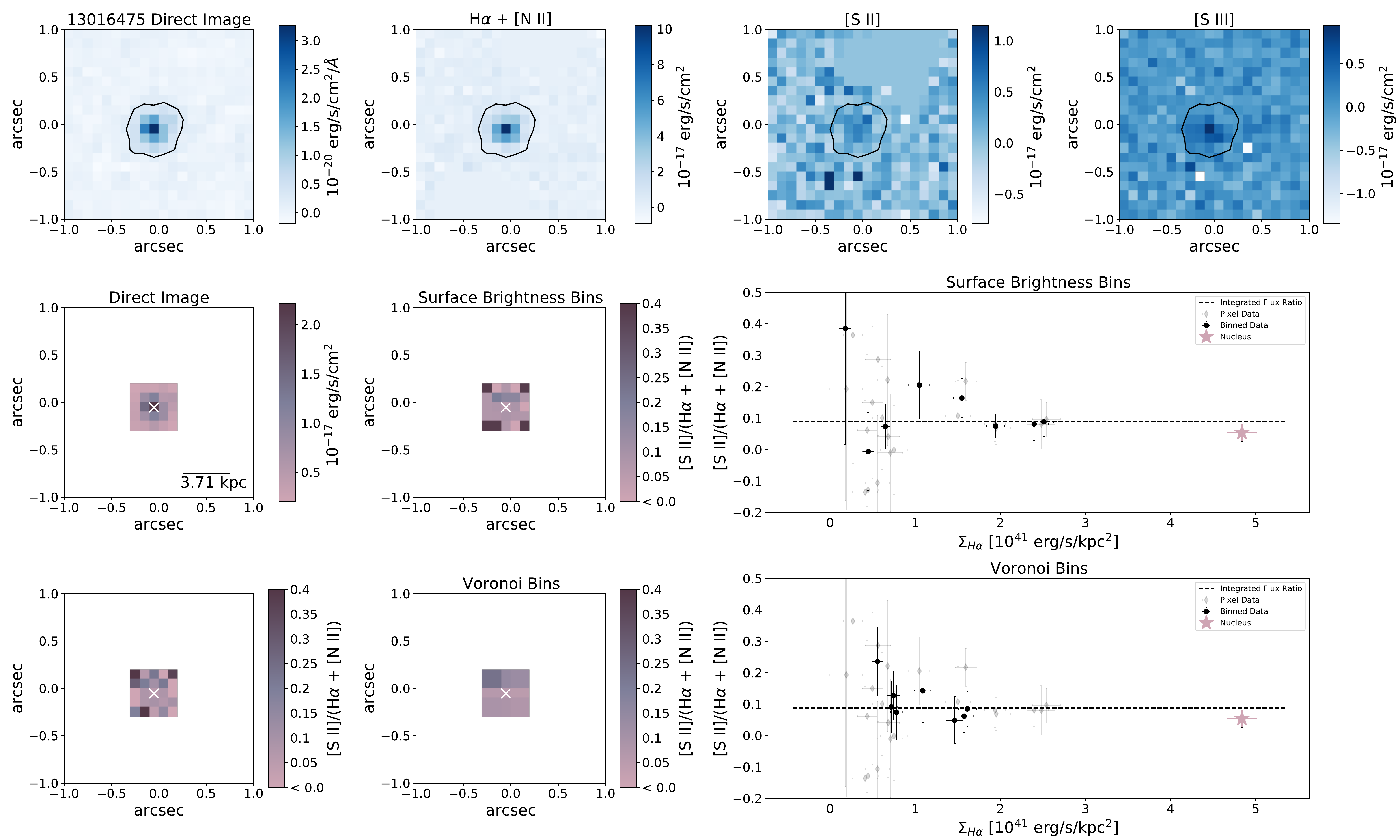}
  \caption{Top Row: Same as Figure \ref{fig:EL_maps}, Bottom Rows: Same as Figure \ref{fig:analysis_bin}}. 
  \label{fig:appendix_first}
\end{figure*}

\begin{figure*}
  \centering
  \includegraphics[width=0.95\textwidth]{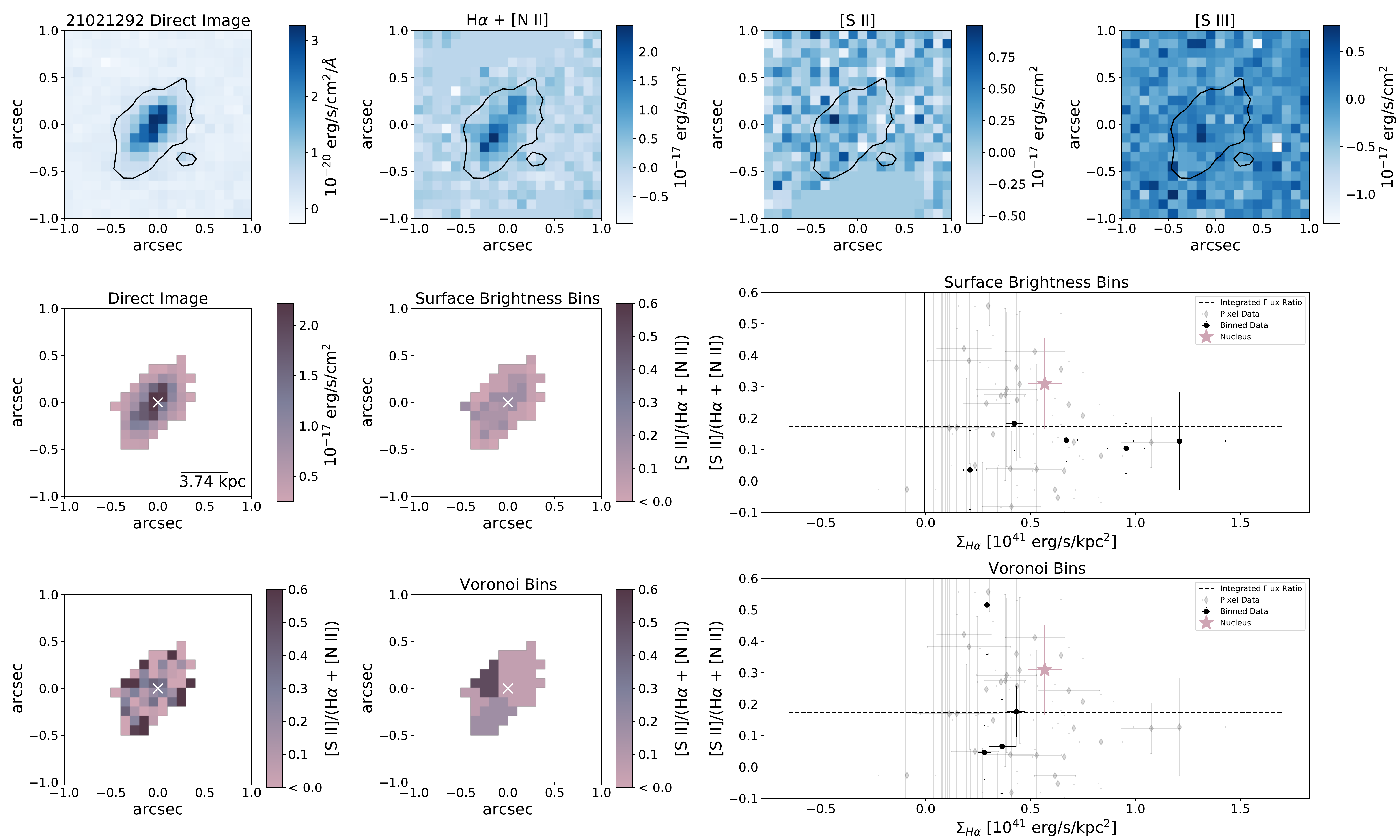}
  \caption{Top Row: Same as Figure \ref{fig:EL_maps}, Bottom Rows: Same as Figure \ref{fig:analysis_bin}.} 
  \label{fig:21021292}
\end{figure*}

\begin{figure*}
  \centering
  \includegraphics[width=0.95\textwidth]{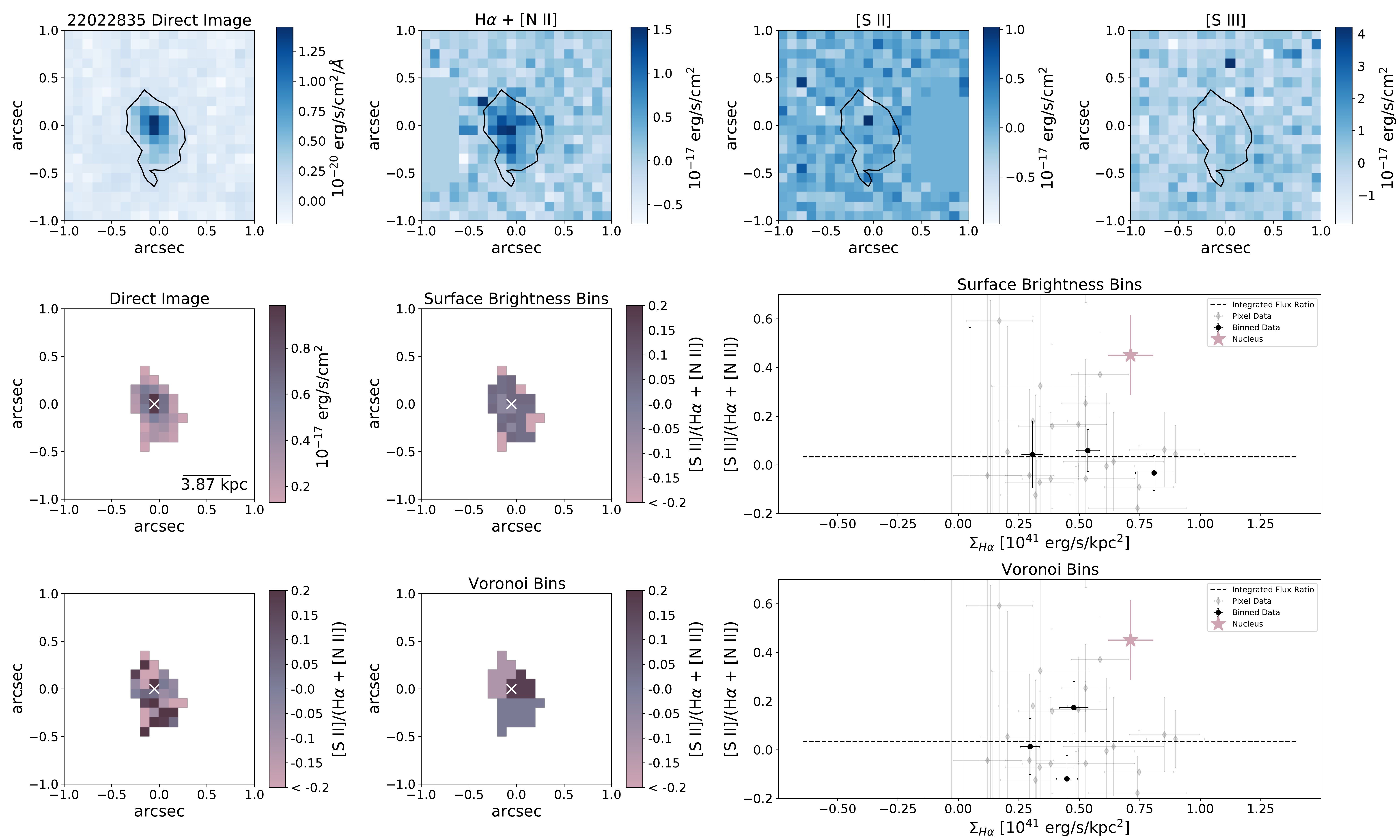}
  \caption{Top Row: Same as Figure \ref{fig:EL_maps}, Bottom Rows: Same as Figure \ref{fig:analysis_bin}.}
  \label{fig:22022835}
\end{figure*}

\begin{figure*}
  \centering
  \includegraphics[width=0.95\textwidth]{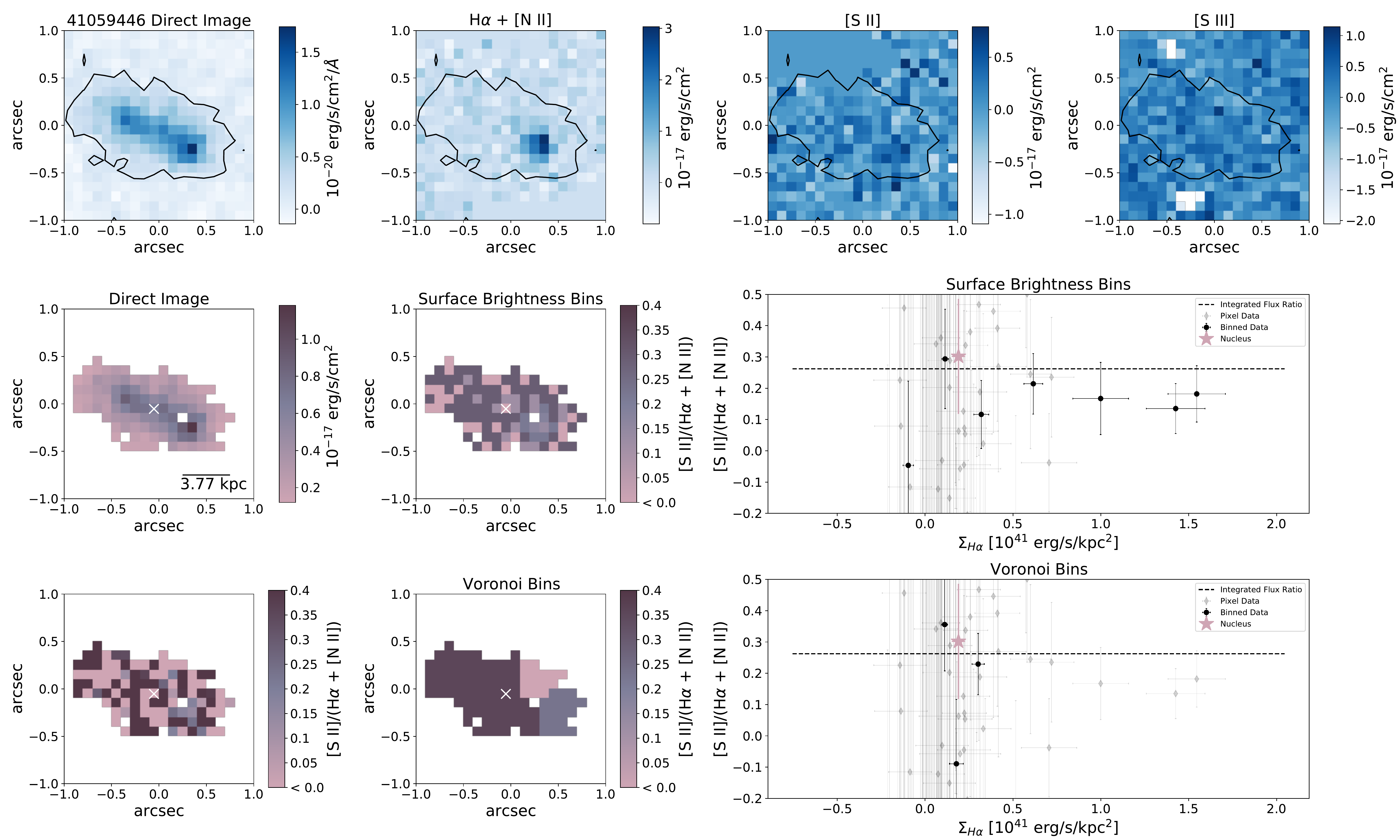}
  \caption{Top Row: Same as Figure \ref{fig:EL_maps}, Bottom Rows: Same as Figure \ref{fig:analysis_bin}.} 
  \label{fig:41059446}
\end{figure*}

\begin{figure*}
  \centering
  \includegraphics[width=0.95\textwidth]{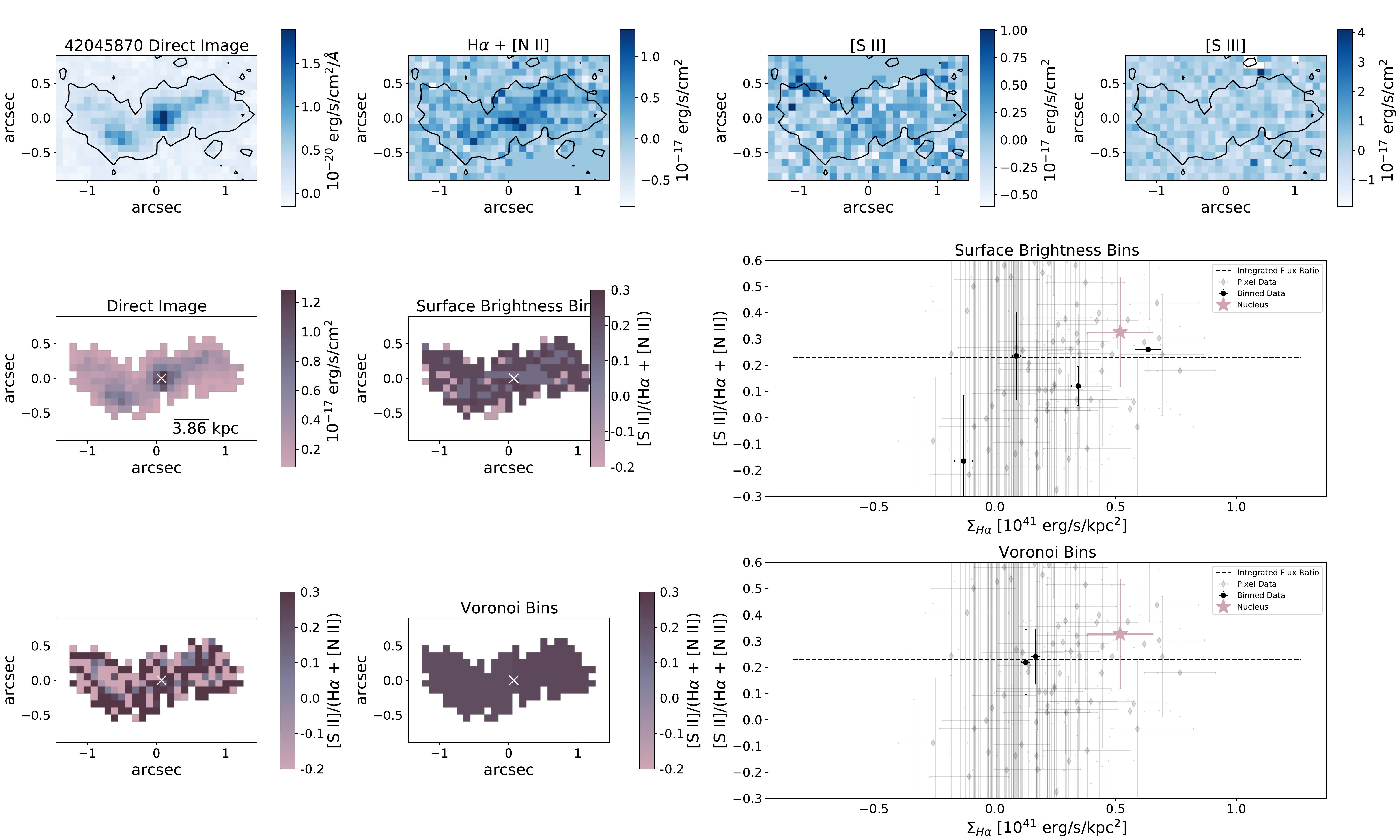}
  \caption{Top Row: Same as Figure \ref{fig:EL_maps}, Bottom Rows: Same as Figure \ref{fig:analysis_bin}.}
  \label{fig:42045870}
\end{figure*}

\begin{figure*}
  \centering
  \includegraphics[width=0.95\textwidth]{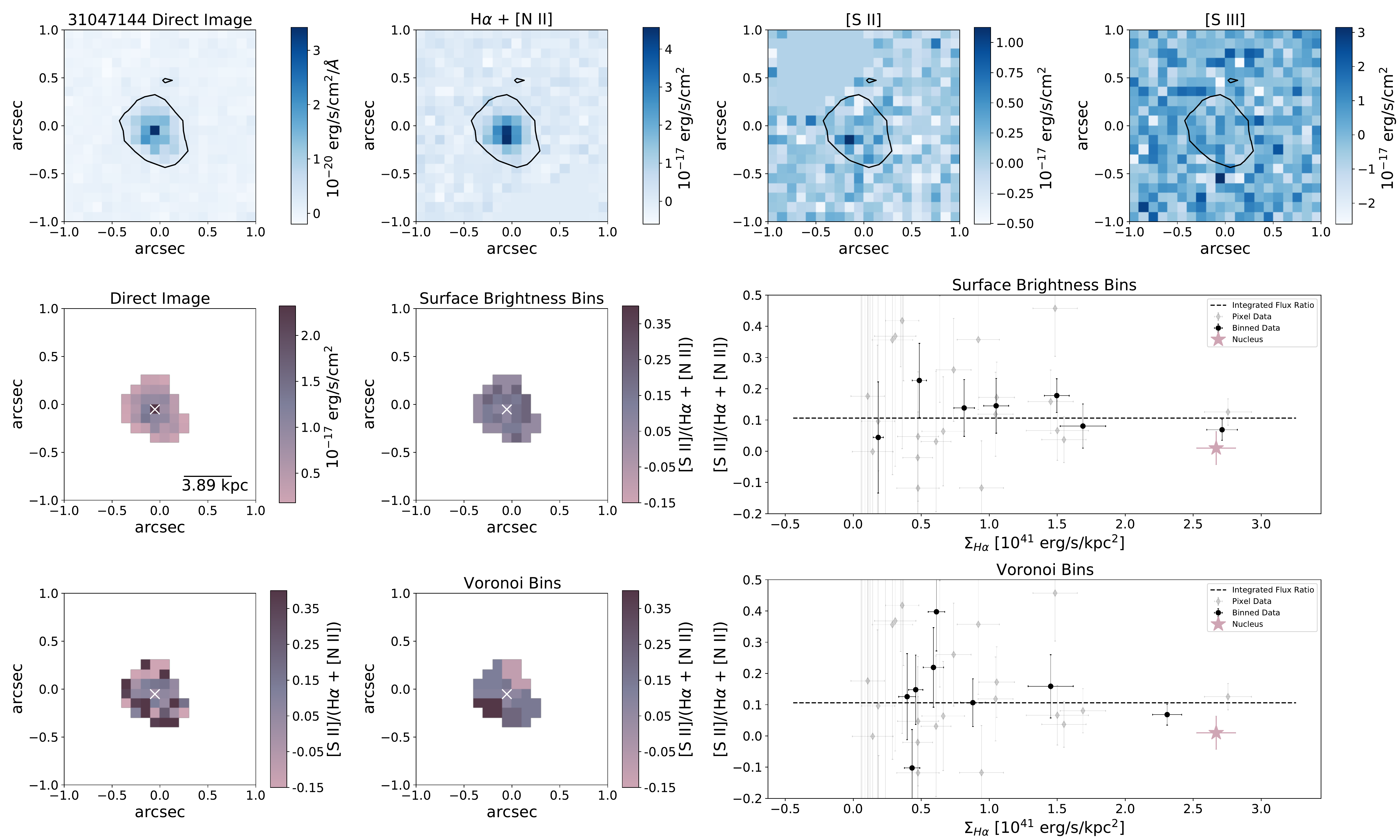}
  \caption{Top Row: Same as Figure \ref{fig:EL_maps}, Bottom Rows: Same as Figure \ref{fig:analysis_bin}.}
  \label{fig:appendix_last}
\end{figure*}

Emission line maps and \SiiHa\ line ratio analysis for the galaxy DEEP2-13043716 are shown in Figure~\ref{fig:analysis_bin}, and interpretation of this example object is discussed in Section~\ref{sec:analysis}. Equivalent data are shown in Figures~\ref{fig:appendix_first}-\ref{fig:appendix_last}.

\begin{figure*}
  \centering
  \includegraphics[width=0.95\textwidth]{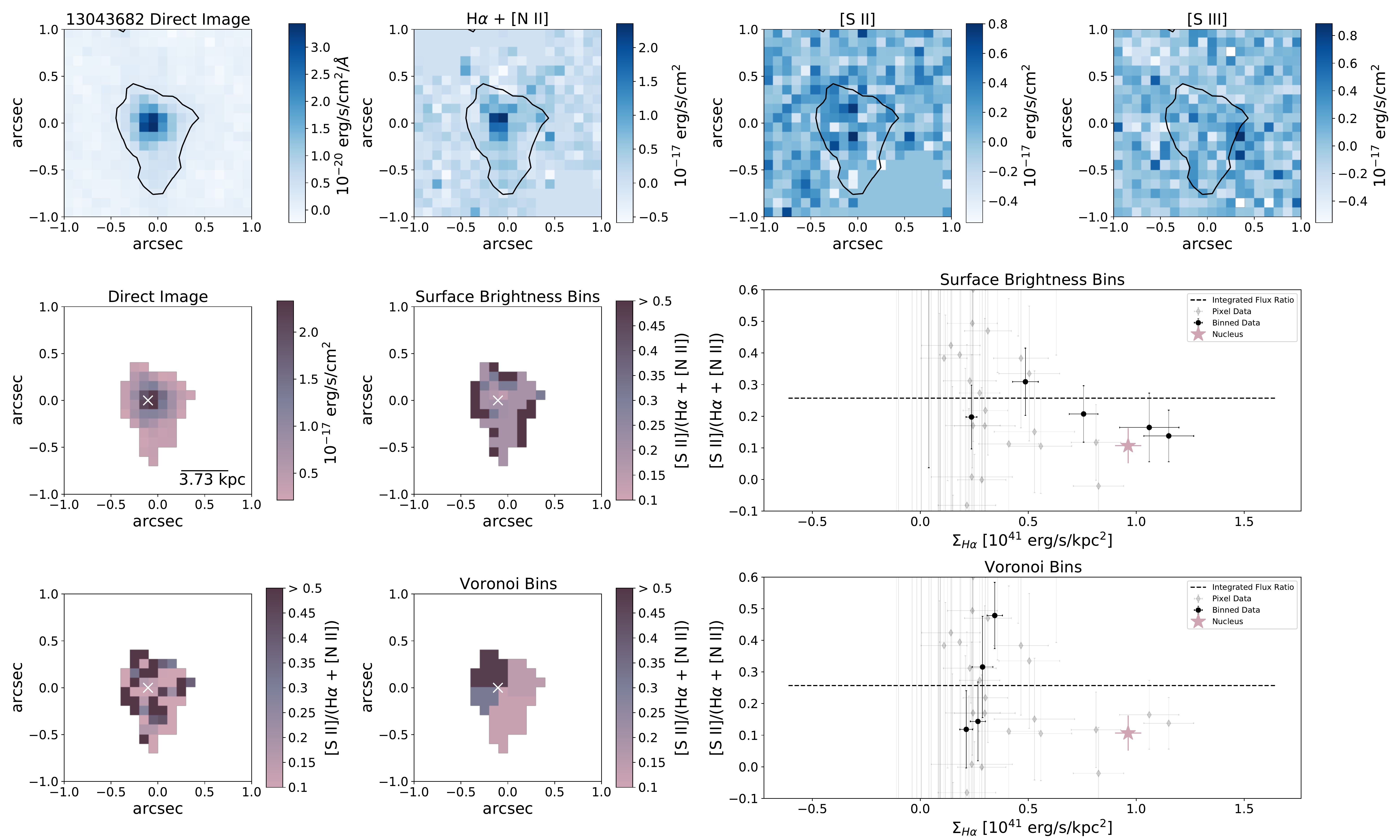}
  \caption{Top Row: Same as Figure \ref{fig:EL_maps}, Bottom Rows: Same as Figure \ref{fig:analysis_bin}.} 
\end{figure*}

\begin{figure*}
  \centering
  \includegraphics[width=0.95\textwidth]{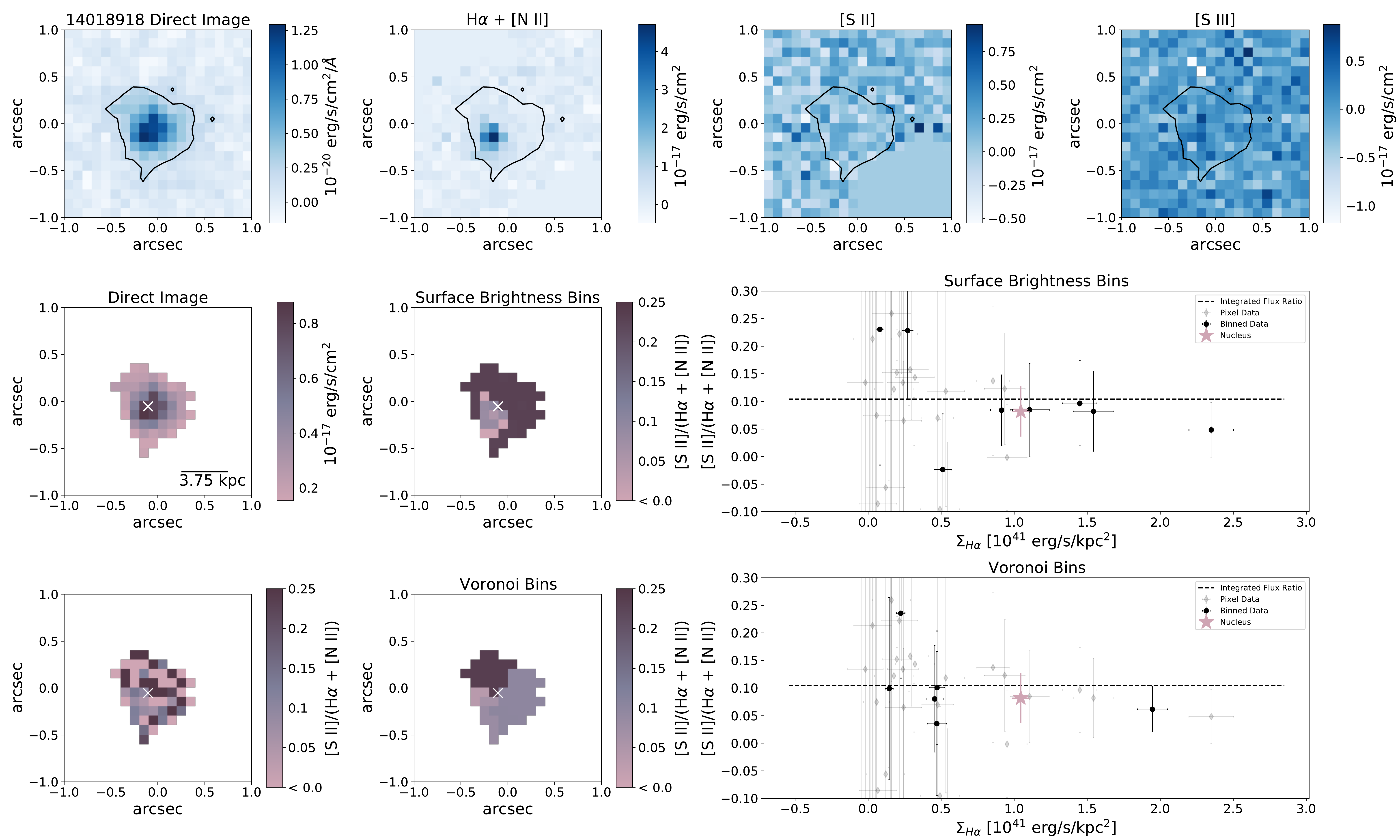}
  \caption{Top Row: Same as Figure \ref{fig:EL_maps}, Bottom Rows: Same as Figure \ref{fig:analysis_bin}.} 
\end{figure*}

\begin{figure*}
  \centering
  \includegraphics[width=0.95\textwidth]{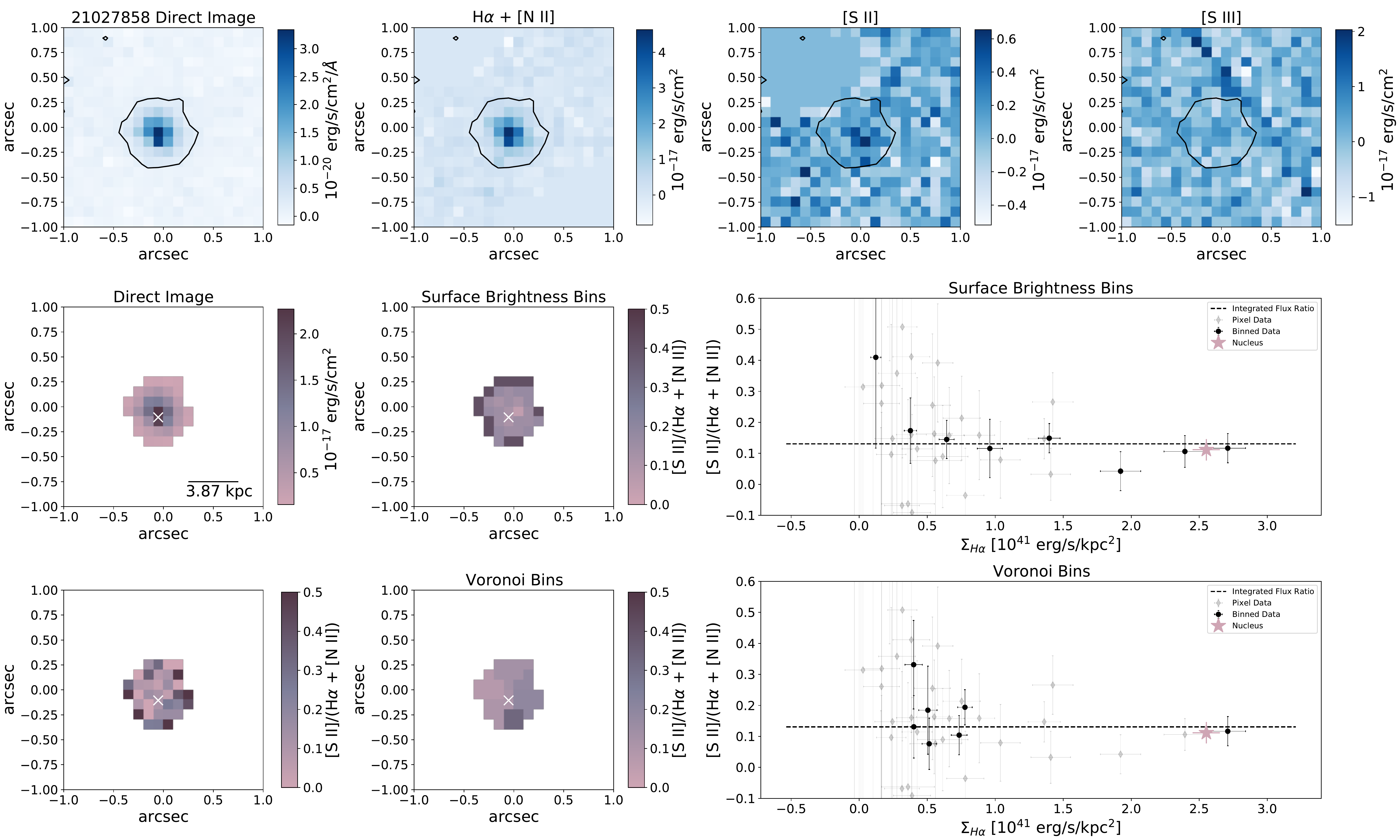}
  \caption{Top Row: Same as Figure \ref{fig:EL_maps}, Bottom Rows: Same as Figure \ref{fig:analysis_bin}.} 
\end{figure*}

\begin{figure*}
  \centering
  \includegraphics[width=0.95\textwidth]{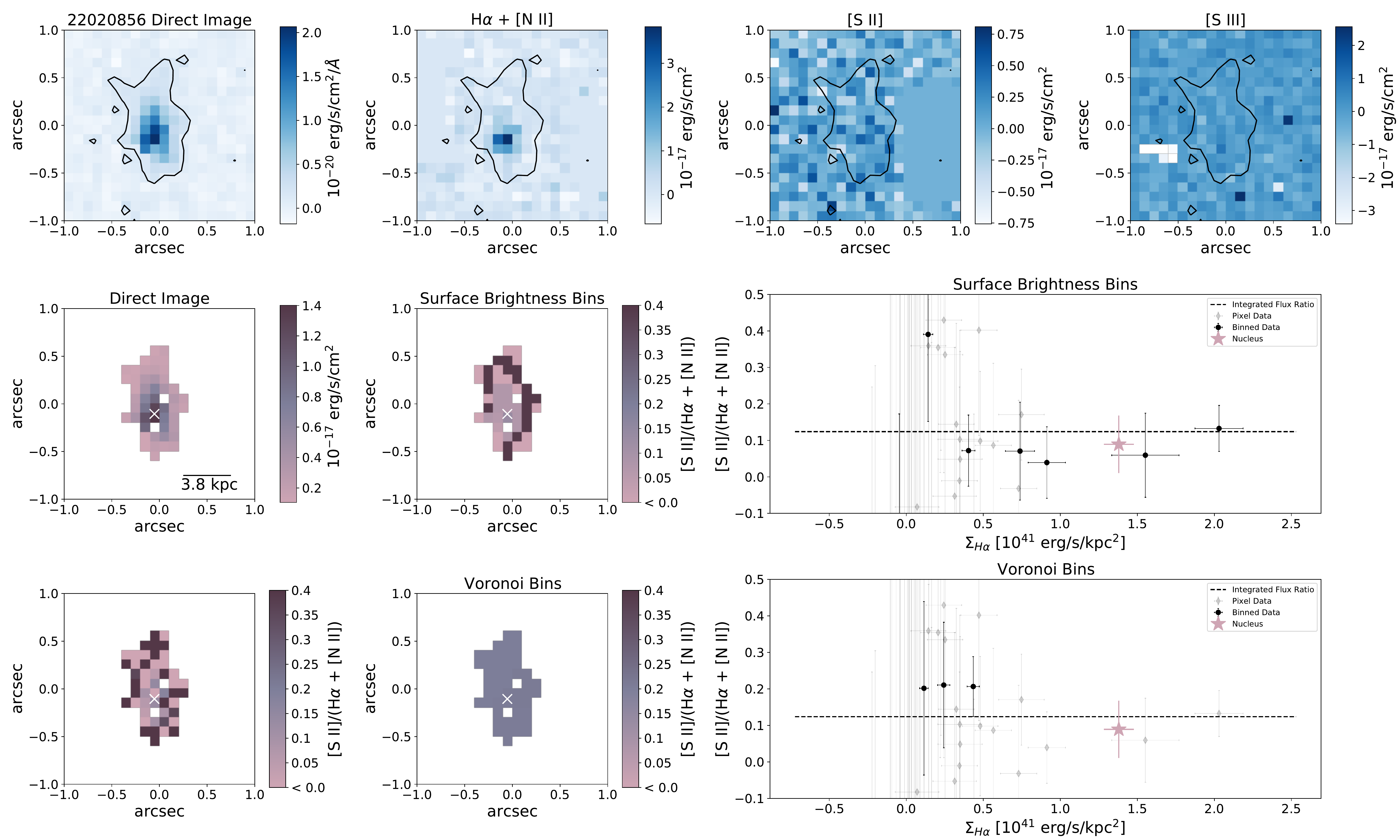}
  \caption{Top Row: Same as Figure \ref{fig:EL_maps}, Bottom Rows: Same as Figure \ref{fig:analysis_bin}.} 
\end{figure*}

\begin{figure*}
  \centering
  \includegraphics[width=0.95\textwidth]{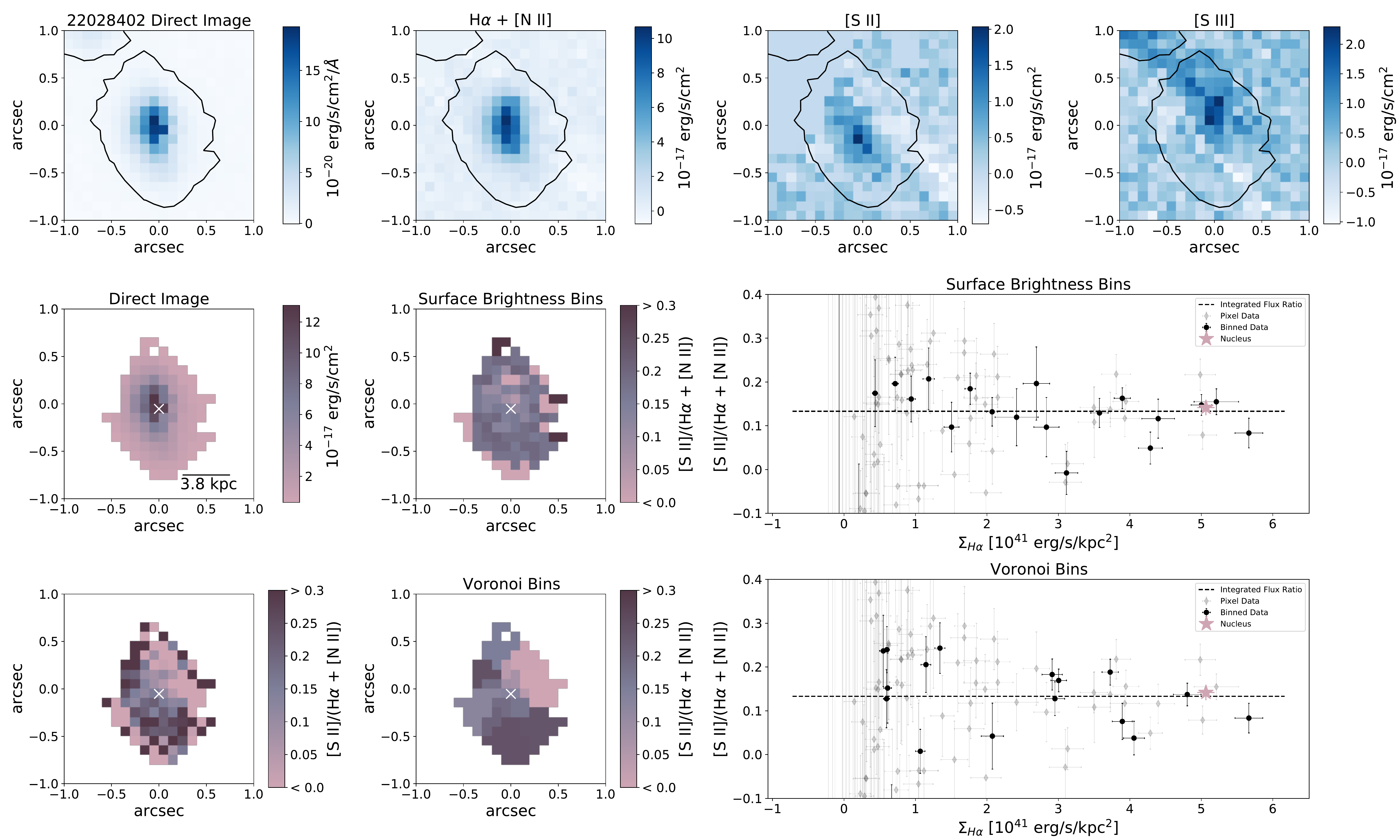}
  \caption{Top Row: Same as Figure \ref{fig:EL_maps}, Bottom Rows: Same as Figure \ref{fig:analysis_bin}.} 
\end{figure*}

\begin{figure*}
  \centering
  \includegraphics[width=0.95\textwidth]{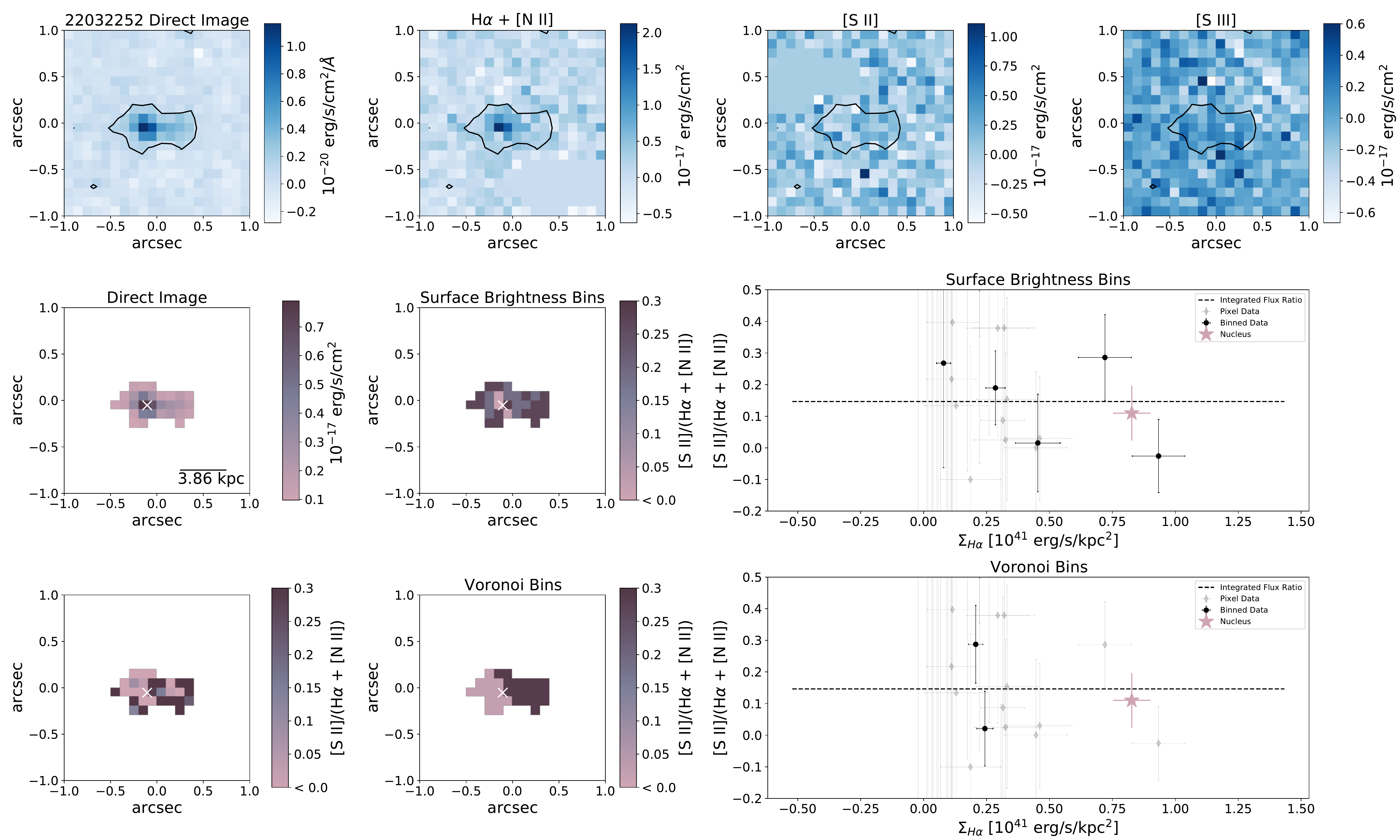}
  \caption{Top Row: Same as Figure \ref{fig:EL_maps}, Bottom Rows: Same as Figure \ref{fig:analysis_bin}.} 
\end{figure*}

\begin{figure*}
  \centering
  \includegraphics[width=0.95\textwidth]{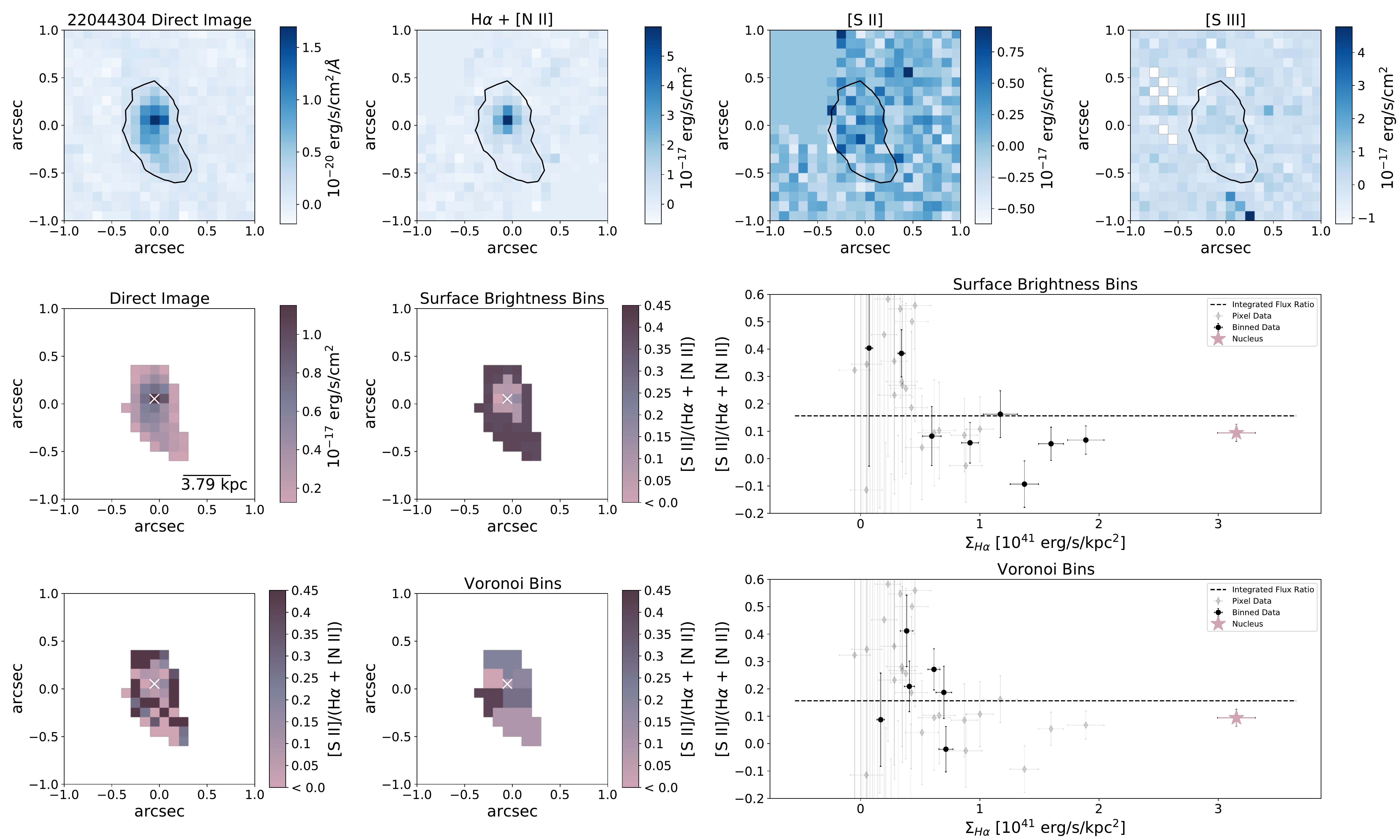}
  \caption{Top Row: Same as Figure \ref{fig:EL_maps}, Bottom Rows: Same as Figure \ref{fig:analysis_bin}.} 
\end{figure*}

\begin{figure*}
  \centering
  \includegraphics[width=0.95\textwidth]{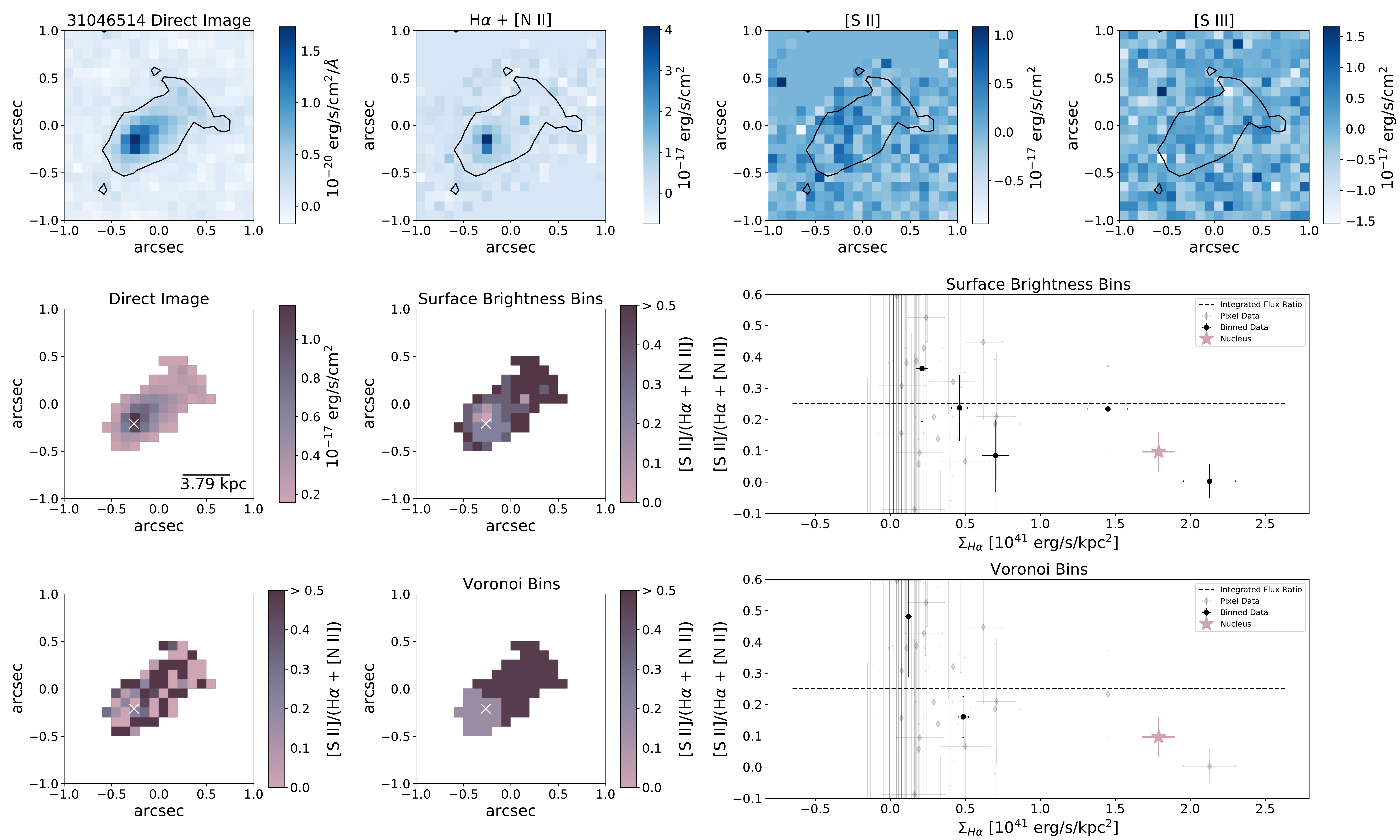}
  \caption{Top Row: Same as Figure \ref{fig:EL_maps}, Bottom Rows: Same as Figure \ref{fig:analysis_bin}.} 
\end{figure*}

\bibliography{HST_bib}{}
\bibliographystyle{aasjournal}



\end{document}